# Where are we heading with NISQ?

Olivier Ezratty [1]

[1] author of the Understanding quantum technologies book and cofounder of the Quantum Energy Initiative, Paris, France,

olivier@oezratty.net

In 2017, John Preskill defined Noisy Intermediate Scale Quantum (NISQ) computers as an intermediate step on the road to large scale error corrected fault-tolerant quantum computers (FTQC). The NISQ regime corresponds to noisy qubit quantum computers with the potential to solve actual problems of some commercial value faster than conventional supercomputers, or consuming less energy. Now, over five years on, it is a good time to review the situation. While rapid progress is being made with quantum hardware and algorithms, and many recent experimental demonstrations using fewer than 50 qubits, no one has yet successfully implemented a use case matching the original definition of the NISQ regime. This paper investigates the space, fidelity and time resources of various NISQ algorithms and highlights several contradictions between NISQ requirements and actual as well as future quantum hardware capabilities. Crucially, either two-qubit gate errors are still around the 0.1%-1% range (with superconducting qubits) or their number capping under 50 (with trapped ion qubits), which limits experiments to rather small algorithms instances that can easily be classically emulated. It then covers various techniques which could help like qubit fidelities improvements, various breeds of quantum error mitigation methods, analog/digital hybridization, using specific qubit types like multimode photons as well as quantum annealers and analog quantum computers (*aka* quantum simulators or programmable Hamiltonian simulators) which seem closer to delivering useful applications although they have their own mid to longer-term scalability challenges. Given all the constraints of these various solutions, it seems possible to expect some practical use cases for NISQ systems, but with a very narrow window before various scaling issues show up. Turning to the future, a scenario can be envisioned where NISQ will not necessarily be an intermediate step on the road to FTQC. Instead, the two may develop along different paths, due to their different requirements. NISQ requires a hundred or so qubits with gate fidelities well above 99.99% to outperform conventional supercomputers in speed or in energy efficiency, while FTQC accepts lesser gate fidelities, around 99.9%, but requires millions of qubits and very long range entanglement capabilities. This leaves open a key question on the trade-offs that may be necessary to make between qubit scale and qubit fidelities in future quantum computers designs.

## CONTENTS

# I. INTRODUCTION

The NISQ era was first defined by John Preskill in his keynote address at the first Q2B conference from QC Ware in California in December 2017 and laid out in a paper published in Quantum in 2018[1]. He then said that "*Quantum computers with 50-100 qubits may be able to perform tasks which surpass the capabilities of today's classical digital computers, but noise in quantum gates will limit the size of quantum circuits that can be executed reliably [...]. I made up a word: NISQ. This stands for Noisy Intermediate-Scale Quantum. Here "intermediate scale" refers to the size of quantum computers which will be available in the next few years, with a number of qubits ranging from 50 to a few hundred [...]. With these noisy devices we don't expect to be able to execute a circuit that contains many more than about 1000 gates*". We have a definition for hardware with over 50 qubits to obtain some potential space related quantum advantage vs classical computers and shallow algorithms that are tolerant to the noise generated during qubit initialization, qubit gates and qubit measurement.

John Preskill added that, and beyond NISQ, "*quantum technology might be preferred even if classical supercomputers run faster, if for example the quantum hardware has lower cost and lower power consumption*". This last part has not been much investigated so far. Most scientific papers published on NISQ algorithms are dealing with some form of computational advantage but not with other kinds of advantages that are more economical in nature, and particularly pertaining to their energetic footprint. Indeed, work must be done to find situations where NISQ systems may someday generate similar results than best-in-class supercomputers or HPCs algorithms, not necessarily faster but, with a lower energy consumption.

## NISQ algorithms classes

The best known quantum algorithms suitable for NISQ systems belong to the broad variational quantum algorithms (VQA) class[2][3]. Given existing and near future qubit gate fidelities, these algorithms quantum circuits should have a shallow depth, meaning a small number of qubit gate cycles, and preferably under 10. This class includes VQE (variational quantum eigensolver[4][5]) for quantum physics simulations, QAOA (quantum approximate optimization algorithm[6]) for various optimizations tasks, VQLS (variational quantum linear solvers[7]) to solve linear equations and QML (quantum machine learning) for various machine learning and deep learning taks. Many other species of NISQ VQA algorithms are also proposed, particularly in chemical simulations[8][9][10][11] and for search[12].

These are most of the time heuristic algorithms that determine near-optimal solutions to various forms of optimization problems, VQE, QAOA and QML being all various breeds of optimization problems to find energy or cost function minima. Variational algorithms are hybrid by design with a very significant part being implemented in a classical computer, a part that is itself a NP-hard class problem that scales exponentially with the input size[13]. Some other non-variational NISQ algorithms are also proposed like quantum walks[14].

Totally outside the NISQ relevant algorithms class are integer and discrete log factoring algorithms (the most known ones coming from Peter Shor in 1994), oracle based search algorithms (like Grover[15] and Simon algorithms), and all algorithms relying on a quantum Fourier transform, including HHL for linear algebra and many partial derivative equations (PDE) solver algorithms. All these algorithms require a fault-tolerant quantum computing (FTQC) architecture, noticeably since, given a number of qubits, typical FTQC gate-based algorithms have a computing depth that grows up on a quasi-polynomial scale with the number of qubits.

In the space and speed domains, a quantum advantage requires at least from 50 to 100 physical qubits. The space and speed domains advantages are however distinct. There are situations where some speedup could be obtained with qubits in the 30-50 range, at least when comparing a QPU with perfect qubits, fast gates and a classical server cluster executing the same code in emulation mode[16], which is usually not the best-in-class equivalent classical solution. Under 18 qubits, it is even recommended to use a local quantum code emulator[17]. It is not only cheaper, but faster and convenient since your computing job is not placed on a potentially long waiting list and you do not have to pay for expensive cloud QPU (quantum processing unit) resources access. A laptop, a single cloud server or server cluster is always cheaper than a quantum computer in that case. As a reference, we propose a taxonomy of various quantum advantages in Figure 29, page 33 in this paper, including space, speed, quality, energetic and cost.

Thus far, most NISQ experiments have been run with fewer than 30 qubits and should therefore better be labelled as "pre-NISQ". While they are elegant proofs of concepts, they do not yet demonstrate any speed up over classical computing, meaning they are not yet in the NISQ regime as defined by John Preskill and listed in Figure 1.

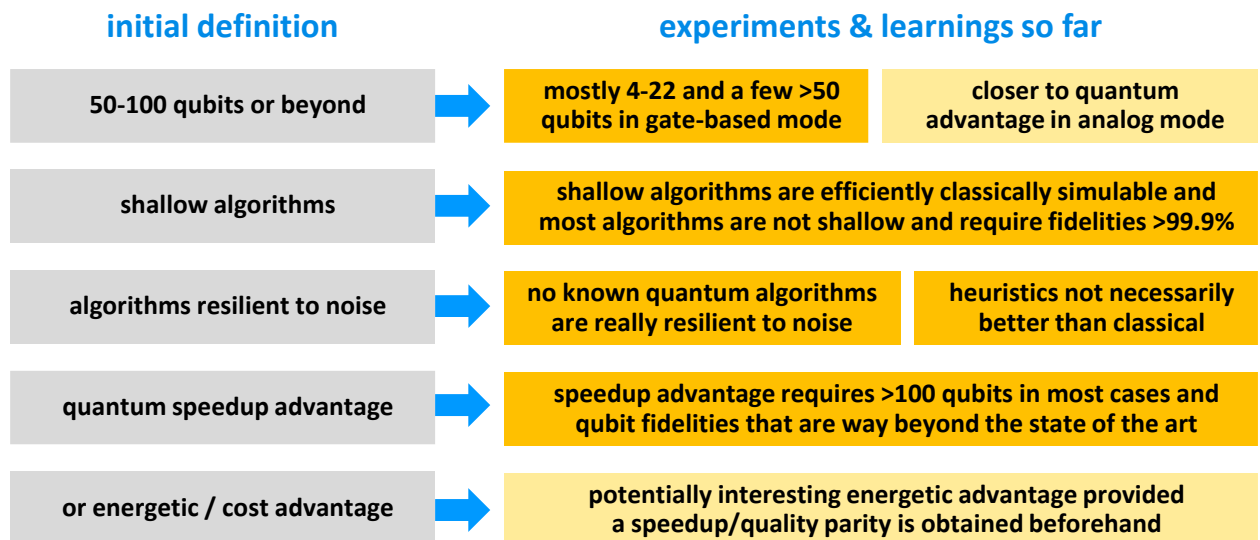

Figure 1: from John Preskill's NISQ definition to actual experiments and learnings. Analog computing refers to digital annealing (D-Wave) and quantum simulators (Pasqal, QuEra, …). Shallow algorithms have only a few gate cycles, preferably under 10. Most variational algorithms have a much larger number of gate cycles in a potential quantum advantage speedup regime. Source: (cc) Olivier Ezratty, 2023.

The aim of these experiments is mainly to verify that a small scale noisy quantum computer can generate some useful results compared to a classical computing system emulating perfect qubits. They are not yet proof of a quantum advantage at a larger scale. Another concern is that is very hard to identify the best-in-class classical solutions which makes it difficult to create apple-to-apple comparisons and well documented quantum speedup assessments, particularly given classical and quantum algorithm don't yield similar results, like a full solution in classical computing versus a value of interest in its quantum equivalent.

When trying to obtain some quantum speedup advantage, existing variational algorithms breadth and depth seem too large for existing NISQ qubit qualities and architectures[18]. There is some hope that quantum error suppression and mitigation techniques may alleviate this requirement but not on a very large scale.

On the other hand, noisy qubits and shallow algorithms can be efficiently emulated with tensor network-based techniques. It can be done efficiently, which means "at most in polynomial time", but not necessarily faster than a quantum computer. And there are only a few benchmarks yet done in that regime[19].

The doubts about NISQ's viability are not fringe in the quantum computing ecosystem. First, there is an ambient criticism of quantum computer vendors who seem to pursue the qubit count quest without taking enough care of their fidelities. I would say that they care about it but currently mostly fail to improve these fidelities for fundamental reasons, but are still making some progress, although not significant enough yet to render NISQ systems commercially viable. Second, there seems to be a relative shift of attention towards FTQC and quantum error correction codes in both academic research and with many quantum computing industry vendors.

The qubit fidelities requirements for useful NISQ and FTQC are different, and their roadmaps can be both interdependent and independent. FTQC could well succeed before NISQ does, given the fidelities required to implement error correction seem less demanding than the fidelities needed for NISQ in the quantum advantage regime. But FTQC faces daunting scaling challenges with enabling large scale and long distance entanglement between myriads of qubits. In other scenarios, NISQ is still positioned as a path to FTQC given they share many common scalability challenges.

**What are experts saying about NISQ?**

Quantum computing vendors and their ecosystems (analysts, service providers, some software vendors) are touting the advent of "quantum computing for business", meaning that their systems are ready for prime time usage[20]. The Q2B conference organized by QC Ware in the Silicon Valley, Tokyo and Paris is about "practical quantum computing". An epidemy of such "quantum business" conferences around the world are in practice overselling NISQ enterprise readiness and urging corporations to jump in the quantum computing bandwagon.

Vendors have an interest to push a story of readiness for quantum computing, at least to attract investors as they are raising funds, and potential customers to drive some revenue which in turn helps get funding. They oversell various use cases which, when you look at the details, represent most of the time to solutions that could be deployed at a much lower cost and even run faster on classical computers, often, even on a simple \$1K laptop. This is a bit different with analog quantum computing solutions which are closer to reaching some quantum computational and economic advantage but don't benefit from the same market push, at least due to the small number of vendors in that space (D-Wave, Pasqal, QuEra).

Some industry vendors like Microsoft, Alice&Bob, QCI, Amazon Web Services (AWS) and PsiQuantum have a story focused on directly targeting the creation of fault-tolerant quantum computers, skipping the NISQ route.

Scientists are split between cautious optimism and plain pessimism. Take for example Daniel Gottesman from the University of Maryland who provided some insights in the 2022 Quantum Threat Timeline Report from the Global Risk Institute[21]. For him, "*It is not clear that there will be any useful NISQ algorithms at all: A lot of the algorithms that have been proposed are heuristic and may not work at all when scaled up. The ones that are not heuristic, like noisy quantum simulations, may not produce useful information in the presence of real device noise. I think there is a good chance \*something\* will work and be useful, but it is definitely not certain.*". In the same report, Shengyu Zhang from Tencent said: "*Most NISQ papers sweep too many issues under the rug, and many don't even show the cost trend with problem size*". Nicolas Menicucci from the RMIT University in Melbourne, Australia, states: "*I don't see NISQ as promising at all. To date, everything useful that a NISQ processor can do can also be done faster on a classical computer. But that pessimism shouldn't be relied upon since it's merely "proof by lack of imagination"*". Indeed, scientists always leave the door open to new discoveries which could change the landscape.

In a February 2023 review paper on superconducting qubits[22], Göran Wendin of Chalmers University bluntly stated: "*Useful NISQ digital quantum advantage - mission impossible? The short answer is: yes, unfortunately probably mission impossible in the NISQ era*".

According to Joe Fitzsimons from Horizon Quantum Computing, "*The hope was that these computers could be used well before you did any error correction, but the emphasis is shifting away from that*"[23]. He even went as far in January 2023 to state in his 2023 predictions that NISQ would simply die, as shown in his tweet in Figure 2.

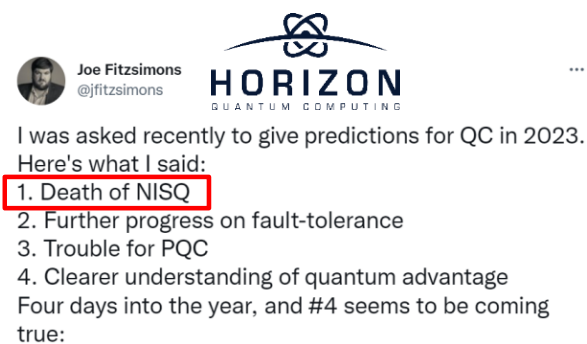

Translated in plain language, it would mean that no quantum computer will be useful until their fault-tolerant breeds are available and running at a sufficient scale, and we're in for at least a decade of waiting. We'll see here that there are a few reasons to be hopeful that, for some computing paradigms like quantum annealing and quantum simulations and some NISQ gate-based algorithms, value can be extracted in the near term for practical usage.

Figure 2: Joe Fitzsimons's provocative quantum forecast post for 2023 on Twitter in January 2023[24]. PQC stands for "post-quantum cryptography", classical cryptographic systems that are designed to be resilient to potential attack of future quantum computers using algorithms like Shor integer factoring algorithm. It is not related to the topic of this paper.

In the remainder of this paper, we'll first look at how to evaluate the hardware and time resources to run NISQ algorithms, then review the state of the art of NISQ variational algorithms which dominate the NISQ scene, and at last, inventory some techniques that are investigated to render NISQ viable, before FTQC shows up progressively.

## II. NISQ COMPUTING RESOURCES

Hardware resource and time estimation is a key quantum computing discipline. There is even a "QRE workshop" for it[25]. It creates a bridge between practical use cases, their related algorithms and their required physical resources and computing time. Late 2022, Microsoft released a resource estimator software tool that does for fault-tolerant quantum computing algorithms[26].

No such generic tool seems to exist for NISQ quantum computing with regards to the number of quantum circuit to run to obtain the expected value of the observable on a given ansatz, the number of ansatzes to run and the cost of the classical part of variational algorithms as described in details in Figure 9[27]. There are however some very interesting review papers documenting well this aspect, at least for VQE algorithms[28 84].

At the same time, any estimation of NISQ resources should be compared to an estimation of the classical computing resources required to solve the same problem. At present, there is a lack of estimators for such best-in-class classical algorithms computing resources. This is always done on a case-by-case analysis, and comparing things with a moving classical target, often in different circumstances, with or without heuristic approaches.

Making a "business" decision of using a quantum computer to solve a given problem would indeed be better off if based on some quantification of its economic cost and benefit compared to existing classical solutions. In classical computing, the "total cost of ownership" (TCO) notion is frequently used but is not yet adopted with quantum computing due to the lack of maturity of the technology and the absence of practical use cases. TCO includes not only hardware and software costs, but also services, training and various direct and indirect solution lifecycle costs. Looking at the current NISQ literature provides, however, some clues.

### NISQ qubit requirements

We will look here at the qubit resources requirements to run NISQ algorithms successfully. Surprisingly, it is not that hard to evaluate. One general rule of thumb determines these physical resource requirements. It links the physical qubit error rate, and the breadth and depth of a given algorithm[29]. The considered error rate corresponds to the gates having the lowest fidelity, which, for most qubit technologies, are two-qubit gates like a CNOT[30].

$$\text{qubit gates error rate} \ll \frac{1}{\text{algo breadth} * \text{algo depth}}$$

The breadth corresponds to the number of qubits used in the algorithms and its depth, to the number of quantum gate cycles. It is a sort of quantum algorithm quantum volume when looking at your quantum circuit. You could make some trade-offs here between these two dimensions and run either a very shallow algorithm with more qubits or a deeper algorithm with fewer qubits. This qubit error rate must be below the inverse of the computing breadth x depth as shown in the above formula[31].

When you compute these numbers with existing quantum hardware, you discover that things don't add up very well, as highlighted in Figure 3. On one hand, to obtain some quantum advantage and match NISQ constraints, you'd need at least 50 physical qubits. On the other hand, the shallowest algorithms have a depth of 8 quantum gate cycles. You end up in that very lower bound case with needing physical qubit gate fidelities of over 99.7%, applicable mainly to two qubit gates and also qubit readout. Today, no single available QPU has such two-qubit gate fidelities with over 50 qubits. Google Sycamore "2022 edition" with 72 qubits has two qubit gates fidelities of 99.4%[32]. IBM's 2020 Prague/Egret system is closer to this threshold with 99.66% fidelities obtained with 33 qubits. IBM expects to reach 99.9% two-qubit gate fidelities with its future Heron 133 qubit processors to be unleashed in 2023. Looking at all vendor roadmaps, IBM is the only vendor expecting to exceed 99% qubit fidelities with over 100 qubits, and possibly even 99.99%. As another example, as shown in Figure 4, Rigetti plans to create a 84 qubits QPU with only 99% two-qubit gate fidelities and, later, a 336 qubits version barely reaching 99.5% fidelities, which is clearly insufficient for running any NISQ algorithm with that number of qubits.

Most two-qubit gate fidelities provided by industry vendors are median or average fidelities. An usually unreported important metric is their standard deviation and minimum values[33]. Good median fidelities with high standard deviation are not at all practical, particularly for the first gates of a given algorithm. High error rates can irreversibly damage early on most running algorithm[34]. One solution consists, after calibration, to deactivate the adjacent qubits for which hardware defects create "stable" faulty two qubit gates. Still, even with using these average fidelities values, the publicized two-qubit gate fidelities are still not good enough to run NISQ algorithms successfully. It is also the case with ion-trap qubits which have very good fidelities but are seemingly hard to scale beyond a couple dozen qubits preventing developers to obtain a space-related computing advantage. These qubits are also too slow to drive, damaging their potential to generate a speedup in a quantum advantage regime[35]. This doesn't show up with most experiments that are implemented with fewer than 25 qubits and a few gate cycles.

# NISQ gate-based hardware resource requirements

| resources | initial estimates | realistic estimates and constraints |
|---|---|---|
| **qubit number** | **50 qubits for a computational advantage (Preskill)** | **100s to 1000s qubits for many practical NISQ algorithms to obtain a speedup advantage (Guerreschi, Albino).** |
| **computing depth** | **use shallow algorithms with under 10-gate cycles** | **most NISQ algorithms in the quantum advantage regime have >100s gate cycles** |
| **available fidelities** | **NISQ is to use currently available qubit fidelities that are in the 99.9% to 99% range** | **current QPUs either have low fidelities and >30 qubits (transmons) or better fidelities and <30 qubits (trapped ions)** |
| **required fidelities** | error rate $\ll \dfrac{1}{\#\ qubits \ast algo\ depth}$  *for QAOA, but seemingly for other NISQ algorithms as well* https://iopscience.iop.org/article/10.1088/2058-9565/abae7d *error rate usually relates to the two-qubit error rate, which should ideally be its minimum error rate and not median/average rate.* | **the fidelities requirements are not matched by actual hardware even for the shallowest computing depth**  1/(1121 q * 8 d) => 99,99%   possible? 1/(127 q  * 8 d) => 99,99%   IBM Heron's 133 qubit QPU in 2024? 1/(65 q   * 8 d) => 99,8%   not available. 1/(53 q   * 8 d) => 99,7%   Google Sycamore is at 98,6%.  ⬑ *minimum ansatz depth of 8 gate cycles* |
| **qubit gates set** | **variational algorithms use single qubit gates $R_x$, $R_y$, $R_z$ with arbitrary angles, on top of Hadamard and CNOT gates** | **arbitrary rotations gates require very high precision and finely tuned calibration, it is a key difference with FTQC which relies on T and Toffoli gates which can be transversally corrected.** |

Figure 3 : table showing the qubits requirements for NISQ algorithms as estimated and with realistic estimates and constraints. It is showing some inconsistencies between the need for more than 50 qubits and their required two-qubit gate fidelities with even the shallowest algorithms. Even with the shallowest NISQ algorithms in the quantum advantage regime, the required fidelities are way above the current state of the art and its expected evolution in the next next few years. On top of that, 8-depth cycles algorithms are easy to emulate with tensor networks on classical computing systems. Arbitrary rotation single qubit gates are another difference with NISQ, which at first glance is an advantage for NISQ implementation as compared to the large overhead of arbitrary rotation gates generations based on fault-tolerant sets including gates like T and Toffoli gates. Source: (cc) Olivier Ezratty, 2023.

In the current industry vendor plans and roadmaps, most QPUs are not expected to scale-in beyond a couple of hundred qubits with supporting the over 99.9% two-qubit gate fidelities required for either NISQ or FTQC.

The most frequently retained option is a scale-out approach with connecting several QPUs together, a bit like with distributed and parallel computing used in high-performance computing (HPC). These connections must preserve qubits overall entanglement and fidelities. Only a few quantum computing companies have started to work on this next stage which could be explored in a parallel way to the development of their QPU. Scale-out architectures can use multiple techniques like microwave guides between qubits or entangled photons-based connections. Specialized quantum information network startups like WelinQ (France) and QPhoX (The Netherlands) have started to build quantum links based on entangled photon-based connections, also providing quantum memories capabilities for computing and intermediate communication buffers.

With several hundred or a thousand qubits, you end up needing gate fidelities between 99.9% and 99.9999% which are clearly out of reach for today's quantum computers even in lab settings with a few qubits as shown in Figure 5[36]. And this ignores the fact that many NISQ algorithms requiring such many qubits are not necessarily as shallow as those requiring only fewer than 10 gate cycles.

There is another notable difference between NISQ and FTQC hardware architectures. As we'll see later, NISQ variational algorithms make a lot of use of R gates with arbitrary rotation angles around the X, Y and Z Bloch sphere axis in their "ansatz" that are prepared classically. These R arbitrary rotations gates must be implemented with very high precision, which is constrained, among other aspects, by the quality of their electronic drive[37] [38].

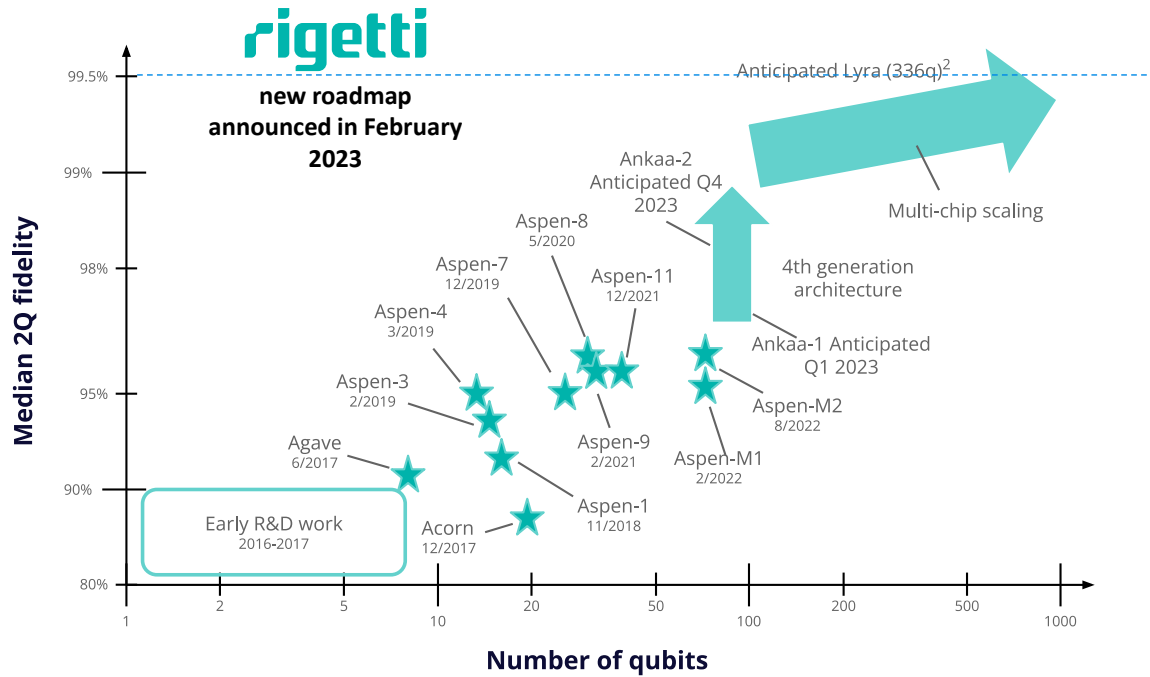

Figure 4: Rigetti roadmap. Source: Rigetti Investor presentation, February 2023[39]. Increasing the number of qubits above 100 while staying below 99.5% fidelities will unfortunately not enable any quantum advantage in the NISQ regime and seems a dead-end path that can't bring any business quantum computing advantage.

With FTQC, these gates are avoided since it is hard to correct their errors in a fault-tolerant manner. They are replaced by a universal gate set usually containing Clifford group gates (Pauli X, Y and Z gates, Hadamard gate, a CNOT gate for entanglement) and a gate enabling the generation of all rotations in the famous Bloch sphere representing the state of a single qubit, usually a T gate (Z rotation with an angle of 45%) or a Toffoli 3-qubit gate.

Then, arbitrary rotation gates are constructed with long assemblies of these primary gates, depending on the needed angle precision, according to the famous Solovay-Kitaev theorem[40]. These gates are used since they can be error-corrected in a fault-tolerant manner which seems not to be the case for arbitrary rotations gates.

### NISQ computing time

Another resource to estimate is the total NISQ algorithm computing time, including its classical portion. After all, we're looking for some computing speedup, but with reasonable computing times related to our patience. Its scaling must be carefully estimated in the quantum advantage regime due to various costs: the number of Pauli strings, the sought precision and the exponential cost of quantum error mitigation, as shown in Figure 9, later in this paper.

Whatever the use cases and speedup, NISQ computing times should be reasonable. We will see that it's not necessarily the case in a quantum advantage regime, when it fares better than classical computing. Most NISQ variational algorithms have a computing time with a lot of variables, equal to $N_i * I_t$ with $I_t = (C_t + S * Q_t)$, with:

$N_i$ = **number of iterations** of the variational algorithm to converge on an acceptable value. It is case dependent and depends on the way the variational algorithm converges to the expected solution.

$I_t$ = **iteration time** to classically prepare an ansatz[41] ($C_t$) and to run it on the quantum computer ($S * Q_t$), representing one iteration, $Q_t$ being the time to run a single shot. $C_t$ also contains the time it takes to handle the classical post-processing of the data coming from the quantum computing shots to generate the expectation value of the Hamiltonian observable from the ansatz. It is highly dependent on the number of shots described below.

$S$ = **number of quantum circuit shots** corresponding to the number of times the ansatz must be executed on the quantum computer to compute the expected value of the observable of the ansatz in order to reach a given precision. This number of shots can scale as high as $O(N^4/\epsilon^2)$, N being the number of useful data qubits and $\epsilon$ the target error rate, with typical VQE algorithms to find the ground state of a molecule[54].

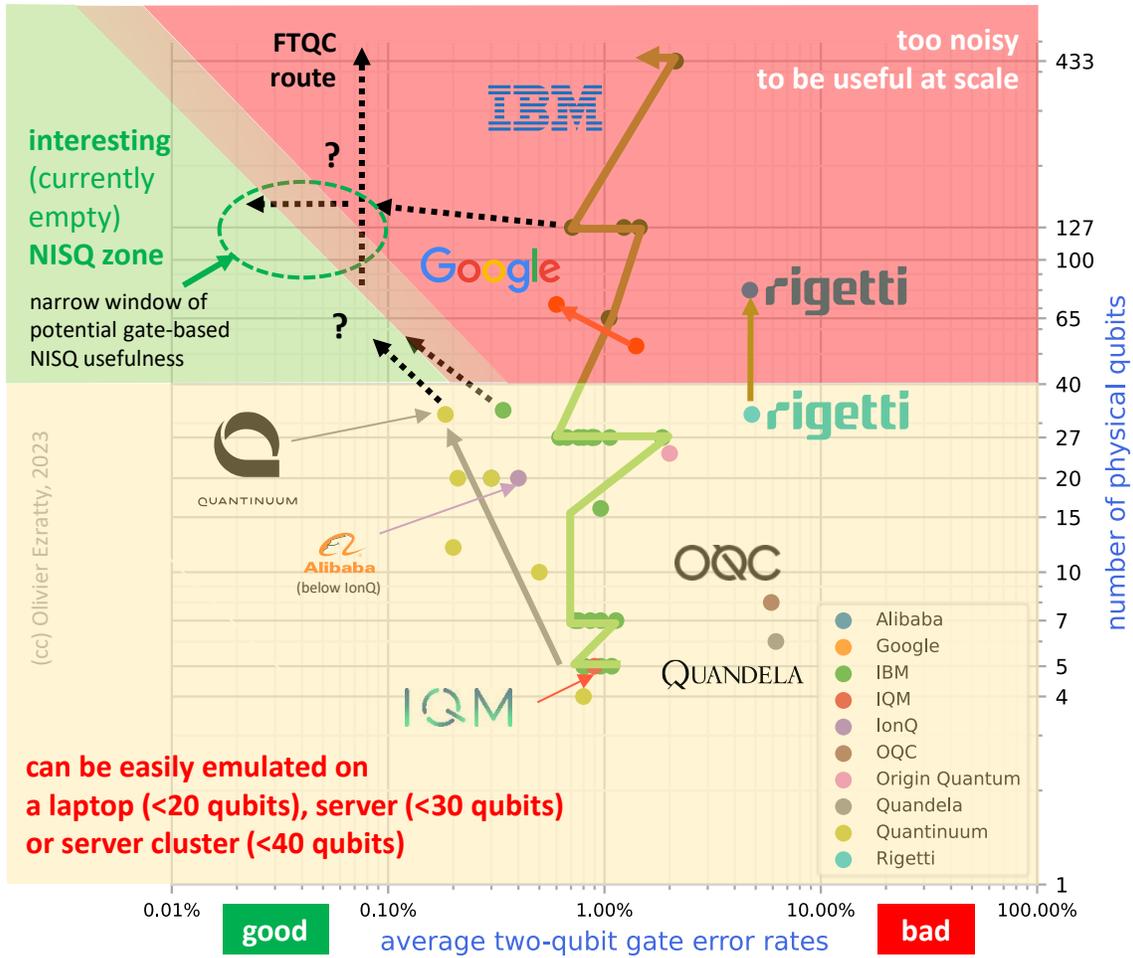

Figure 5: scatter plot with two-qubit gate fidelity and qubit number for currently available commercial gate-based quantum computing systems. Viable NISQ QPUs require figures of merit that are in the empty slanted upper-left green zone. It is slanted since, as the qubit number grows, qubit fidelities must be better to accommodate a larger quantum volume. IBM's zigzag corresponds to continuous fidelities improvement within each QPU generation having several iterations. The yellow zone corresponds to a quantum computing regime that can be easily emulated with the demanding "state vector" mode on classical computers. It is faster and cheaper under 20 qubits with a simple laptop, faster with an SV1 AWS server instance under 29 qubits and possible with a cluster server like the Eviden (Atos)[42] QLM under 40 qubits with a 1 TB memory, and up to 44 qubits with AWS cloud servers[43]. Two other figures of merit are missing here like qubit connectivity, which impacts algorithms depth and quantum volume which describes the usefull breadth (number of qubits) and depth (actual algorithm gate cycles) on these systems. Source: Olivier Ezratty, Kordzanganeh et al [17] and vendors two-qubit gate fidelities data obtained with randomized benchmarking[44], and plotted over QPUs number of qubits in log scales, as of May 12th, 2023. Qubit fidelities data correspond to average fidelities, regardless of their standard deviation. With large standard deviations, actual NISQ algorithms are significantly damaged by gate errors. See Tannu and Qureshi[33].

The $O(N^4)$ scale corresponds to the number of Pauli strings, which applies a basis change with series of single qubit gates, changing the computational basis before qubits readouts[5]. It is like doing a partial state tomography of the data qubits[45]. For example, determining the ground state of a molecule such as benzene with its 12 atoms ($C_6H_6$), would require 72 qubits and running 330,816 Pauli strings in a VQE algorithm[46][47]. Some Pauli strings can however be regrouped with some preprocessing, reducing the number of Pauli strings to $O(N)$, but at the cost of longer circuits that may be prohibitive in NISQ regime[84]. With QAOA algorithms, the number of Pauli strings scales as $O(2L)$[48], L being the number of rotations/entanglement cycles in the quantum circuit. Measurements are done for each Pauli string with $O(1/\epsilon^2)$ circuit executions shots[49], corresponding to an outcome precision $1 - \epsilon$, which by the way does not correspond to the chemical accuracy obtained as a result, that is also damaged by the qubit noise[50].

Then, some classical postprocessing generates the ansatz objective function result[70]. Some other tactics are proposed to optimize the number of required shots[51], including at the error mitigation stages[52].

The target error rate $\epsilon$ can be very low for VQE algorithms used in quantum chemical simulations, increasing as a result the number of required shots to astronomical levels. In 2015, it was estimated that finding the ground energy state of ferredoxin ($Fe_2S_2$) with 112 spin-orbitals with VQE would require $10^{19}$ circuit shots and $10^{26}$ gate operations[53]. Various optimizations are proposed to remove the polynomial or exponential curse against the number of data qubits and they are algorithm dependent[54][55]. It can otherwise become a key showstopper of NISQ implementations beyond N=40, and to reach some practical quantum advantage.

You can then complement this list with the various quantum error mitigation techniques overhead which further increases the number of shots and adds some more classical processing burden. This overhead scales exponentially with the circuit depth or qubit number depending on the used mitigation techniques. With qubits having sufficient fidelities, making rather simple chemical computations with optimized VQE algorithms could last several decades if not centuries with superconducting qubits[84]. How about the better trapped ions qubits with their high fidelities? These qubits are completely out of the game here, due to their quantum gates that can be about 1000 times slower than with superconducting qubits[56]. A theoretical speedup compared to classical computing is of no value if it practically happens at non-human time scales!

Again, a practical full-stack evaluation of all these time costs would be useful when discussing potential NISQ quantum advantages. It is not always studied in many NISQ algorithms papers which mostly deal with sub-NISQ scaling regimes with fewer than 30 qubits. It still drives some interesting architecture designs where many of these numerous shots would be run in parallel either on different QPUs or even, within a single QPU that would be logically divided in several small qubit zones running the same circuit[57].

**NISQ code classical emulation**

There are two main ways to assess the differences between quantum computers and classical computers. A simpler and imperfect one is to compare a given quantum algorithm execution on a QPU and its code emulation on various types of classical computers. This emulation can be achieved by reproducing the behavior of perfect qubits (with state-vector emulation) or of noisy qubits (using density matrices or the tensor networks technique). The other is to make a similar comparison, but with a best-in-class classical algorithm serving the same need as the quantum algorithm. Indeed, a best-in-class classical algorithm may be faster than the quantum algorithm simple emulation on a classical setup. Comparisons between classical computers and NISQ systems must also consider the various subtleties related to heuristics, output sampling, finding one solution vs finding the best solution and the likes.

All these comparisons are properly done in only a few cases. We are left with guessing what type of NISQ quantum algorithm could be emulated or not on a classical system and to compare their relative speed, cost and energy spent. On top of this, emulation is not a one-stop-shop solution since it can be implemented in various ways, emulating perfect qubits, using for example state vectors, or handling some compression techniques like with using tensor networks that can work with a large number of qubits with shallow algorithms and are relevant for NISQ code emulation, as shown in Figure 7 with data from Nvidia. But we don't need to be too picky with the details. Some key thresholds can be defined between different levels of quantum code emulation based on the number of qubits and the algorithm depth, as shown in Figure 6.

Also, a "quantum advantage" usually shows-up when a QPU has at least the same capability of the most powerful supercomputers, but this equivalence can be assessed when doing a comparison with regular less powerful HPCs. In that case, would the QPU sizing be much different? Would the classical solution be less expensive than the quantum one? How much? This is an open question. In the NISQ regime, things get complicated since all quantum algorithms are hybrid and require a significant classical part to prepare the "ansatz" that is repetitively adapted and run on the quantum processor. In the case of QML algorithms, the classical computer does a lot of work with data ingestion and preparation like doing some vector encoding for natural language processing tasks. In the case of a comparison with some classical code emulation, the classical emulator should be paired with the same classical computer handling the classical part of the algorithm.

One classical way to gauge classical emulation capacities is to assess the memory available. The capacity doubles for each additional qubit to emulate. In state vector emulation mode, which is the most demanding, 29 qubits require 8 GB of memory which fits well in most laptops nowadays[58].

But there are some differences between memory and processing requirements. A powerful laptop with 16 GB of memory may not be sufficient to emulate 29 qubits faster than a QPU. An Amazon SV1 cloud instance is faster than a QPU with that same number of qubits[17].

One Intel server node can emulate up to 32 qubits[59]. While an Eviden (Atos) QLM can emulate up to 40 qubits with over 1 TB of RAM, the related execution time may be longer than on a QPU, regardless of the results quality. GPU-based emulation is the most efficient one so far, with Nvidia leading the pack with its series of V100, A100 and the most recent H100 GPGPUs, their general purpose GPUs that serve different needs than the GPUs used in gaming and 3D images rendering[60].

All this is well summarized in the above chart in Figure 5 that shows the connection between qubit numbers, two-qubit gate error rates and various ranges for their emulation. The chart reminds us that no single vendor has yet developed a QPU in the "useful" zone of >50 physical qubits and >99.9% qubit fidelities. Note that IBM recently improved its two-qubit gates fidelities with using ECR gates (echoed cross resonance gates) that are different from a CNOT[61]. Other similar plots have been created like the one coming from the Unitary Fund METRIQ initiative, against time of benchmark instead of number of qubits and using more rigorous benchmarking techniques independent from the vendors and using QPUs available in the cloud. The fidelities are slightly different, but the picture is about the same[62].

| Number of qubits | NISQ hardware availability | Resources for classical emulation in state vector mode or MPS mode | Emulation mode | Classical emulation computing depth |
|---|---|---|---|---|
| 1 to 18 | Yes | Laptop, faster than QPU. | State vector | Unlimited |
| 18 to 30 | Yes | Server, faster than QPU. | State vector | Unlimited |
| 31 to 40 | Yes | Server cluster, Eviden (Atos) QLM. | State vector | Unlimited |
| 41 to 55 | Yes | HPC and supercomputers for large depth circuits. | State vector or tensor networks / MPS | Unlimited |
| 56 to * | 127, 433 qubits | Possible with tensor networks and compression techniques on shallow algorithms and noisy qubits using MPS. | Tensor networks / MPS | Limited |

Figure 6: table assessing the typical classical resources needed to emulate a gate-base quantum algorithm. MPS stands for matrix product state, a tensor network-based method used to emulate shallow gate-based quantum algorithms on classical systems. Source: Olivier Ezratty and Xiaosi Xu et al for details on time/memory scaling[63].

Using a tensor network base compression technique like DMRG (density matrix renormalization group), an Eviden (Atos) QLM was used to digitally emulate Google's 53-qubits Sycamore supremacy random sampling in about 30 hours vs less than 3 minutes with Sycamore[64]. According to Thomas Ayral et al[65], there won't be any computational quantum (exponential) advantage with NISQ systems. They did argue that the cost of NISQ code emulation is growing linearly with the number of qubits when their fidelity is under 99.9%.

Sandbox AQ and Google also broke some records in 2023 with using a TPU-v3 pod supercomputer to implement DMRG code, thanks to its fast distributed matrix multiplications capacity, originally built to train large machine learning models. This code is used to compute the ground state of a local quantum many-body Hamiltonian, a classical equivalent of a NISQ solution that would be implemented with a VQE algorithm. In that case, Google could support a bond dimension of $2^{16} = 65,536$, that sizes the number of entanglements in the simulated many-body system[66]. In another work from Honghui Shang et al (China), 1,000 qubits and chemical simulations using a VQE algorithm were digitally emulated on a Sunway supercomputer[67].

Daniel Stilck Franca and Raul Garcia-Patron wrote similarly, in 2020, that "*Noise can make VQE and QAOA algorithms easy to simulate on classical computers*" and precises that "*noise mitigation techniques based on post-processing of the quantum computation measurement outcomes, despite being useful to filter the data from noise, would not change the predictions of our work*"[68].

We are in a situation where NISQ quantum computing advantage cannot be obtained with existing quantum hardware and qubit fidelities, or we can emulate it efficiently on classical computing. This doesn't bode well for NISQ but we'll later see that some solutions loom around to fix some of these issues.

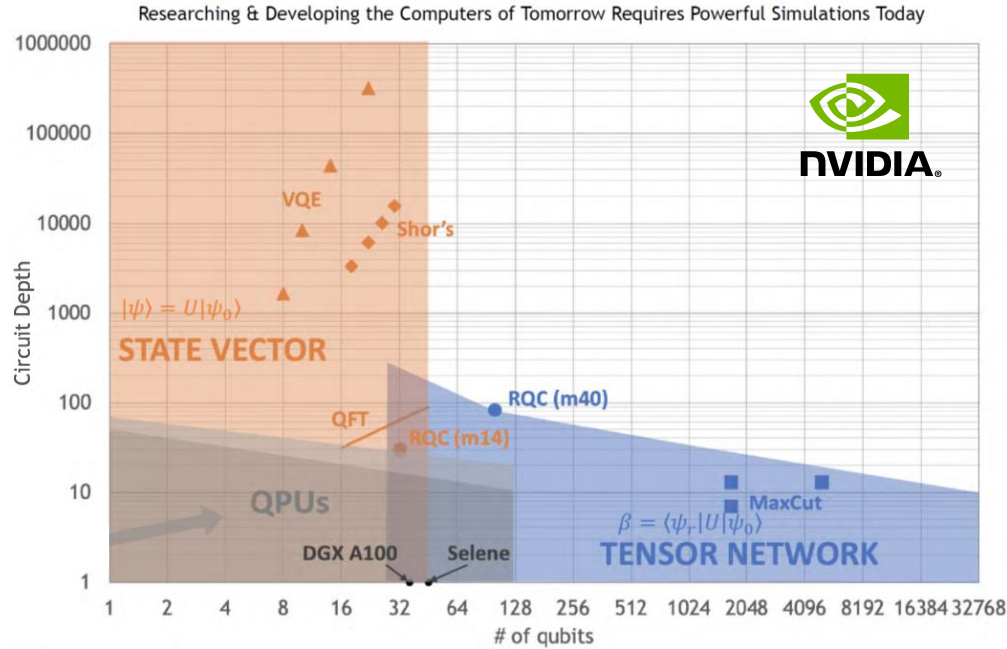

Figure 7: Nvidia positioning the scope of state vector quantum emulation in a regime with fewer than 32 qubits but no limitation in the circuit depth (in the Y axis), and tensor networks emulation which can scale with hundred of qubits with a shallow algorithm. This last solution is adequate to emulate NISQ algorithms with not many limitations. The figure shows that classical emulation has a broader scope than existing NISQ quantum computers (in grey). The Shor point in the scatter plot corresponds to running Shor integer factoring algorithm on very small RSA keys, not the sought-after RSA-2048 key. It is the same with the VQE point which corresponds to rather small chemical simulation requirements. Source: Nvidia[69].

### III. NISQ ALGORITHMS RESOURCES

We will now make a review of the quantum algorithms that are suitable for NISQ QPUs and focus not on their underlying principles but on their qubit resource requirements and computing time scale. Several review papers make good inventories of what could be potentially achieved with NISQ algorithms.

Bharti et al[70] state in their 91 pages NISQ algorithms review that "*These computers are composed of **hundreds of noisy qubits**, i.e. qubits that are not error-corrected, and therefore perform imperfect operations in a limited coherence time. In the search for quantum advantage with these devices, algorithms have been proposed for applications in various disciplines spanning physics, machine learning, quantum chemistry and combinatorial optimization. The goal of such algorithms is to leverage the limited available resources to perform classically challenging tasks.*". It is interesting in the first place that they position NISQ in the hundreds of qubits range.

Jonathan Wei Zhong Lau et al write in another review paper[71] on the state of NISQ that "*NISQ algorithms aim to utilize only shallow-depth quantum circuits (right now, around a few hundred gates in depth at maximum)*" which, considering that 50 qubits (the lower bound definition of NISQ) times 100 gate cycles engender the need for physical qubits two-qubit gate fidelities of 99.98% which is way out of the current hardware capabilities. They note accordingly that "*we may be in this era for a long time*".

As shown in Figure 8, some algorithms are not relevant in the NISQ mode (in orange). Neither the Shor integer factoring nor the Grover search algorithm are appropriate for NISQ since they use either complicated circuits like parametrized period finding and an inverse quantum Fourier transform, or an oracle function.

Jonathan Wei Zhong Lau et al also write that "*A heavier reliance on analog computing as opposed to digital computing might also be necessary*" which also corresponds to our finding detailed later in the "NISQ enablers" part. They also express doubts on the viability of quantum simulations (VQE) and machine learning (QML) algorithms on NISQ platforms. They still expect positive feedback loops as more developers are testing quantum algorithms on existing QPUs.

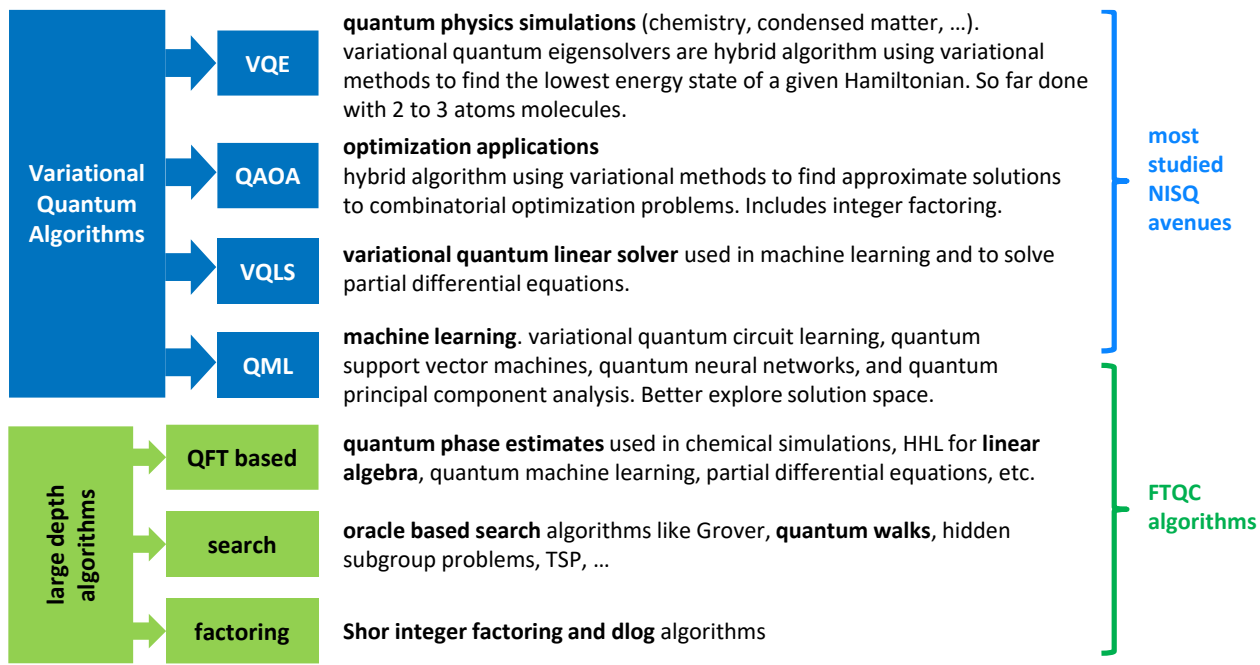

In another work, Chen et al position Grover and shadow tomography algorithms[72] NISQ versions near the classical BPP class (solvable in polynomial time by a classical computer, so providing no quantum speedup) and Bernstein-Vazirani algorithm near the BQP class (class of problems solvable in a polynomial time by a FTQC, so potentially providing some quantum speedup vs classical algorithms) [73]. This work, summarized in Figure 10, although not very favorable to the studied NISQ algorithms, seems to neglect the fact that complexity classes deal with asymptotic computing computational complexity. We've seen that NISQ algorithms can't practically scale well and will therefore probably never reach asymptotical realms.

Practically speaking, the most studied NISQ algorithms belong to the variational quantum algorithms class[74]. It includes mainly VQE for chemical simulations, QAOA for combinatorial optimizations as well as many QML algorithms. All these are hybrid classical-quantum algorithms and heuristics based. They use an ansatz function that computes the Hamiltonian of a quantum system parametrized by many rotations of arbitrary angles of single qubit $R_x$, $R_y$ and $R_z$ gates completed by some CNOT gates. These parameters are initialized and tuned by the classical part of the algorithm as shown in Figure 9. The number of quantum shots and ansatz recomputations depend on the algorithms, its data and the precision sought.

A theoretical proof that a depth-3 quantum circuit on an arbitrary number of qubits cannot be emulated in a polynomial time by a classical algorithm was created in 2004 by Barbara Terhal and David DiVincenzo[75] and completed the same year by Scott Aaronson[76]. But this proof and others does not seem to account for the detrimental noise of NISQ circuits and the cost of error mitigation techniques[77].

Also, most known NISQ algorithms are variational with a part that runs on a classical computer and prepares the ansatz that runs in the quantum computer in a loop fashion until convergence is obtained[78]. Several questions deserve attention here. The first one is how would the classical part scale with large problems in the NISQ quantum advantage regime?

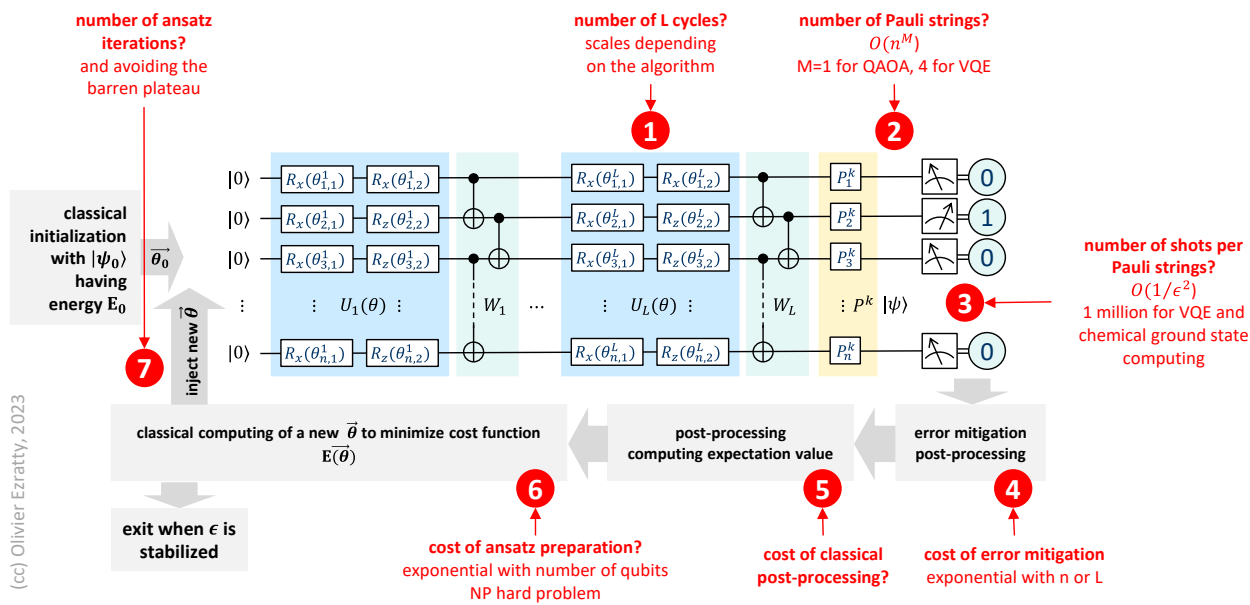

Figure 9: chart describing how variational quantum algorithms (VQA) operate and their scaling parameters. The grey part corresponds to the classical components of these algorithms. An ansatz contains a Hamiltonian encoded with single rotation and two-qubit CNOT gates cycles. It is prepared classically to generate an expected value of the Hamiltonian after computing several runs. Additional ancilla qubits and operations can be added to the ansatz and are not shown here for simplification. The key scaling parameters here are: (1) the number of qubits and of phase and mixing operators in the ansatz (which in that case is labelled a "Quantum Alternating Operator Ansatz", another QAOA, also used in VQEs[79]) determining the circuit depth and conditions the required qubit fidelities, given the computing depth is at least equal to the number of qubits due to the number of entangling gates to execute and the SWAP gates used with nearest neighbor qubits topologies[86], (2) the number of Pauli strings for the measurement of the expected values from observables of the computed Hamiltonian which can scale polynomially with the number of qubits for VQE algorithms[80] but scales linearly with QAOA optimization algorithms, (3) the number of shots to obtain a given precision for each Pauli string which scales as high as $O(1/\epsilon^2)$, meaning one million shots for typical chemical simulation precisions of one per thousand with VQE algorithms, (4) the additional cost of quantum error mitigation which can scale exponentially with the circuit depth or qubit number, (5) the classical post-processing to compute the cost function value, (6) the classical cost to prepare each ansatz, which is usually an NP complete problem, and (7) the number of ansatzes to converge on a satisfying cost function value, avoiding the barren plateau syndrome. The number of quantum circuit shots can become gigantic in the quantum advantage regime, particularly with VQE algorithms which require very high accuracy for chemical simulations. Source: (cc) Olivier Ezratty, 2023 and Jules Tilly et al[84].

## BPP ⊊ NISQ ⊊ BQP

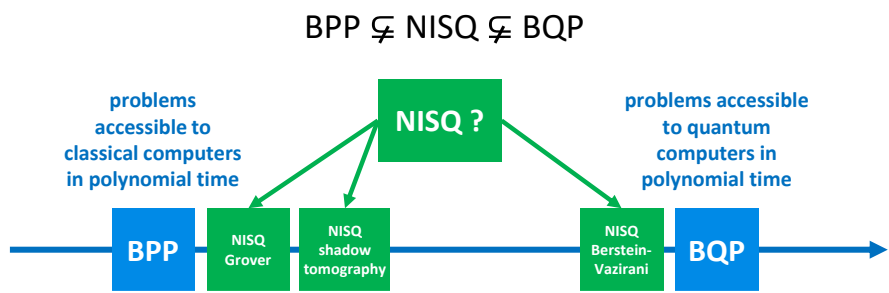

Figure 10: Chen et al study the complexity class of three quantum algorithms in the NISQ regime. They find that Bernstein-Vazirani algorithm is in a class close to BQP, so providing a quantum advantage whereas NISQ Grover and shadow tomography algorithms are near BPP, providing no quantum speedup. But putting NISQ algorithms in a complexity classification is not consistent with complexity classes definitions which deal with asymptotical limits. Due to their characteristics and hardware constraints, noisy quantum algorithms and circuits don't scale well to asymptotical limits.

Kenneth Rudinger from the US Department of Energy Sandia Labs declared that "*the variational approach might not be practical when quantum computers finally become capable of living up to their promise. We have good reason to believe that the size of the kinds of problems you would want to solve is too large for the variational approach; at that scale it becomes essentially impossible for the conventional computer to find good settings for the quantum device*" [81]. To address that problem, a Sandia Labs research team introduced FALQON, a MaxCut QAOA variant performing ansatz optimizations without an expensive classical optimization loop[82]. It converges to good approximation ratios and success probabilities with reasonable resources scaling. However, FALQON uses deeper circuits than classical QAOA which turns into a higher requirement regarding physical qubit fidelities. It's hard to have it all!

Another important and usually unaddressed question is what is the relative weight of the classical computing part in the variational quantum algorithms, in computing time and total cost of operations? At this point, most papers don't elaborate much on the classical resources cost of variational algorithms. The total classical and quantum cost in the quantum advantage regime is supposed to be favorable when compared to a best-in-class full classical algorithm. These best-in-class full classical algorithms should be mentioned in the literature, particularly when dealing with combinatorial optimization algorithms.

### VQE algorithms resources

To date, most VQE experiments were implemented with a few qubits, way under the quantum advantage threshold, nearly always way under the 50 qubits mark. There are several reasons for these experiments being done in a pre-NISQ regime, way below 50 qubits. First, many projects by PhD candidates last between one and three years. Second, while several QPUs are available with over 50 qubits, particularly from IBM and Google, these have qubit gate fidelities too low to enable larger scale VQE (and VQA) noisy-resilient algorithms. The real usable QPUs quantum volumes are very low, with a record of $2^{22}$ obtained with Quantinuum trapped ions QPUs[83]. These experiments are useful to test algorithms whereabouts before QPUs can scale and accommodate a larger number of qubits. As Simone Severini from AWS wrote me in April 2023, "*NISQ hardware is useful in science, but unclear if useful in business*".

These experiments most often deal with condensed matter physics, nuclear physics, high-energy particles physics, vibrational and vibronic spectroscopy, photochemical reaction properties predictions, to name a few, as described in the excellent Tilly et al VQE review paper[84].

In the chemical simulation realm, VQE experiments are usually limited to finding the ground state energy of the Hamiltonian of simple two to three atoms molecules like LiH, $BeH_2$ or $H_2O$[85][86]. As we've seen before, finding the ground state of a slightly more complicated molecule as benzene drives NISQ systems in uncharted territory and very long computing times and requirements for very high-fidelity physical qubits [47][87]. Results show that, for a wide range of molecules, even the best-performing VQE algorithms require gate-error probabilities on the order of $10^{-6}$ to $10^{-4}$ to reach chemical accuracy. VQE can also help compute the excited states of molecules[88][89].

VQE is not yet addressing more pressing computational chemistry needs like determining large molecular structures, finding complex vibrational and rotational spectra, and molecular docking that are all useful in drugs design and in the chemical industry. These use cases belong generally to the FTQC regime, and in most cases, in extreme situations with very large numbers of logical qubits. For example, estimating the ground state of a complex molecule Hamiltonian in the FTQC domain is to be based on the quantum phase estimate (QPE) algorithm. Its precision depends on the number of ancilla qubits in which the eigenvalue result is encoded.

Here's a (certainly incomplete) inventory of some of these relatively recent VQE experiments on real hardware and summarized in the table from Figure 11.

- Ruslan N. Tazhigulov et al from Google[90] used the QITE method (quantum imaginary time evolution) with quantum error mitigation to simulate molecular structures like Fe-S cluster[91] and α-RuCl3 with between **3 and 9 qubits** on a Sycamore processor. Starting at 11 qubits with 1092 gates, their simulation was unsuccessful.

- Arute et al implemented a Hartree-Fock simulation VQE algorithm with error mitigation with **12 qubits** on Sycamore[92]. They state that "*It is still an open question whether NISQ devices will be able to simulate challenging quantum chemistry systems and it is likely that major innovations would be required*".

- Armin Rahmani et al created a 1D setting with a linear circuit depth with the number of qubits, experimented with **12 qubits** running Google's Sycamore QPU[93].

| Qubits # | QPU | Year | Author | Resource estimates in a quantum advantage regime? |
|---|---|---|---|---|
| 27 | IBM | 2023 | Chen et al. | No |
| 3 to 9 | Sycamore | 2022 | Tazhigulov et al. | Unsuccessful with 11 qubits. |
| 16 | IBM | 2022 | Koh et al. | No. |
| 57 | IBM | 2022 | Frey et al. | No. |
| 5 | IBM | 2022 | Kirmani et al. | No. |
| 3 to 9 | Trapped ions | 2022 | Zhu et al. | No. |
| 20 | Sycamore | 2021 | Xiao Mi et al. | No. |
| 9 | Trapped ions | 2021 | Paulson et al. | No. |
| 12 | Sycamore | 2020 | Arute et al, Google AI. | No. |
| 12 | Sycamore | 2020 | Rahmani et al. | No. |
| 6 | IBM | 2019 | Smith et al. | Mentioned the need for better fidelities qubits. |
| 4 | IBM, Rigetti | 2018 | Cervera-Lierta. | To find. |

Figure 11: table summarizing the mentioned NISQ algorithms papers and the number of qubits on which they were tested. It also mentions whether some resource estimates are provided for the extension of these low-NISQ regime implementations to quantum advantage levels. Most noticeably, none of these papers make a comparison with a classical system with regards to any consideration, execution time or cost, whether in emulation mode or with best in-class digital simulations. Source: (cc) Olivier Ezratty, 2023.

- Chen et al on quantum simulation evaluating the ground state of an isotropic quantum Heisenberg spin-1 model on a **27-qubit** IBM QPU[94]. It uses post-selection and ancilla qubits. The paper mentions many NISQ algorithm experiments that helped me inventory many of the other experiments in that list. This list is described like this: "*Programmable digital quantum computers have so far been **successfully used** for the implementation and study of discrete time crystals (DTC), quantum chemistry problems with Hartree-Fock methods, fractional quantum Hall states, spin chain dynamics, interacting topological lattice models, many-body localization, lattice gauge theory and quantum spin liquid states*". Successfully, yes, but always at a small scale.

- Ammar Kirmani et al tested an isolated 1D chain of **5 qubits** with error mitigation techniques[95].

- Adam Smith et al tested a condensed matter time evolution using a Trotter decomposition of the unitary time evolution operator with **6 qubits** on a 20-qubit IBM QPU[96]. They report that their "*benchmark results show that the quality of the current machines is below what is necessary for quantitatively accurate continuous-time dynamics of observables and reachable system sizes are small comparable to exact diagonalization. Despite this, we are successfully able to demonstrate clear qualitative behaviour associated with localization physics and many-body interaction effects*".

- Alba Cervera-Lierta made an exact Ising model simulation with **4 qubits** on IBM and Rigetti QPUs[97].

- Jin Ming Koh et al simulated a quantum topological fermionic system with **16 qubits** on a 27-qubit IBM QPUs[98].

- D. Zhu et al implemented a computation of spectral functions on a trapped-ion quantum computer for a one-dimensional Heisenberg model with disorder with between **3 and 9 qubits** on a trapped ion QPU[99]. The circuit depth contains 96 gate cycles, and it is executed 2,400 times.

- Danny Paulson et al implemented a quantum simulation of 2D Effects in lattice gauge theories on **9 qubits** from a trapped ion QPU[100].

- Xiao Mi et al from Google simulated a discrete time crystal (DTC) on an isolated 1D chain of **20 qubits** in Google Sycamore that is emulable on a laptop[101].

- Philipp Frey et al tested discrete time crystal simulation on a 1D chain of **57 qubits** from a 65-qubit IBM QPUs (now retired). This is the only VQE experiment from this list implemented with over 50 qubits, but without any quantum advantage[102].

- A last interesting example is a paper by Anton Robert et al on an efficient NISQ grade protein folding algorithm[103]. This hybrid protein folding algorithm uses a variational quantum algorithm and a classical genetic algorithm. It scales in polynomial time and computing depth as $O(N^4)$ with N being the number of monomers in a tetrahedral lattice. It was tested in 2019 to fold a 10 amino acid Angiotensin peptide on 22 IBM qubits, which was a best-in-class QPU back then. However, the paper does not mention the performance of the classical DeepMind AlphaFold that was then available in its first version launched in 2018[104] nor results from the various CASP (Critical Assessment of Structure Prediction) programs that serves as a benchmark in this field. Also, since the algorithm depth scales polynomially with the size of the proteins to fold, it quickly exceeds the capacities of noisy qubit systems as the number of amino acids grows. AlphaFold can fold in ternary structures proteins with up to 450 amino acids way above what the NISQ algorithm mentioned above could implement.

VQE is sometimes described as the most appropriate VQA subset of algorithms that are suitable for NISQ QPUs. According to Sebastian Brandhofer et al[105], VQE chemistry simulation algorithms do not scale in the quantum advantage regime unless qubit gate fidelities are very good with error rates below 0.18%[106]. These gate fidelities are not available yet, particularly over 50 qubits, whatever the qubit technology. Other researchers point to chemical simulations requiring very high precision, which is hard to obtain in NISQ regimes, up to a point that FTQC versions of VQE algorithms are proposed[107], but their computing time is totally prohibitive, even for small molecules[108]. At last, a June 2023 preprint from Thibaud Louvet, Thomas Ayral and Xavier Waintal finds that qubit noise prevents NISQ VQE from providing sufficient chemical accuracy in chemical simulations[109].

### QAOA algorithms resources

QAOA is the second most relevant class of VQA algorithms for NISQ QPUs. However, despite it requires fewer shots than VQE algorithms, it seems that it doesn't scale well and requires a larger number of higher quality qubits than are currently available to bring some quantum advantage with practical use case in the enterprise operations domain[110][111].

A QAOA algorithm often relies on a QAOA component. This acronym strangeness comes from the Quantum Alternating Operator Ansätze (QAOA), the ansatz circuit that is used within a variational algorithm, alternating single qubit rotation gates and CNOT gates, as shown in Figure 9[112].

Anton Simen Albino et al state that "*thousands of qubits will be needed before QAOA and its variants can be used to solve these problems, due to the linear relationship between the dimensionality of the problem and the number of qubits. However, the qubits used will not necessarily be error-corrected due to the characteristics of the heuristic itself, which requires low-depth circuits and few measurements of the final state*", in a paper dealing with solving partial derivative equations (PDEs) in fluid mechanics[113].

Johannes Weidenfeller et al provides a lot of clues on QAOA running on NISQ systems[114]. They highlight some obstacles to overcome to "*improve to make QAOA competitive, such as gate fidelity, gate speed, and the large number of shots needed*". Their paper covers transpiler optimizations techniques and how QAOA works with the IBM heavy-hex qubit connectivity. It also provides an estimation of the number of shots to $O(n^2/\epsilon)$, n being the number of qubits and $\epsilon$ the expected algorithm precision. A large number, even if much lower than VQE shot counts.

In a paper dealing with using QAOA to solve a graph partitioning Max-Cut problem, G. G. Guerreschi and A. Y. Matsuura conclude that "*quantum speedup will not be attainable, at least for a representative combinatorial problem, until several hundreds of qubits are available*"[115]. In their work, they make a classical comparison using a single Intel Xeon Phi processor. Such a single CPU would beat a QPU until it reaches about 900 qubits. 900 qubits and even a shallow algorithm would indeed land us in the high-fidelity qubit requirement territory zone with 1/(900*8) error rate, so 99.9986% (see Figure 12)[116]. Meanwhile, most QAOA experiments are done with only a few qubits[117][118]. A Max-Cut problem may be even more demanding in precision than a VQE used for some chemical simulation[119]. Various tricks are proposed to reduce the circuit depth of QAOA ansatzes and are slightly moving the needle in the direction of a potential quantum advantage[120][121][122][123][124], some being hardware dependent[125][126].

Guillermo González-García et al land with the same conclusion[127]: "*We find that, even with a small noise rate, the quality of the obtained optima implies that a single-qubit error rate of 1/(nD) (where n is the number of qubits and D is the circuit depth) is needed for the possibility of a quantum advantage [...]. We estimate that this translates to an error rate lower than $10^{-6}$ using the QAOA for classical optimization problems with two-dimensional circuits*". And with 1000 qubits (see Figure 13)!

As a direct consequence, FTQC and over a million physical qubits seem to be required for implementing QAOA algorithms in the quantum advantage regime! One workaround would be to build relatively large scale NISQ systems with high qubit connectivity, a topic we'll investigate later in the "NISQ enablers" section.

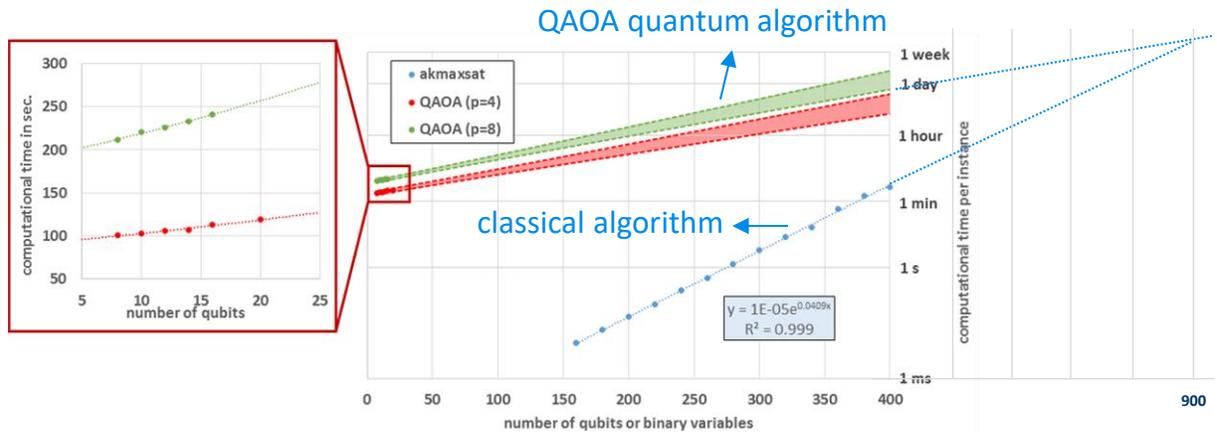

Figure 12 : QAOA hardware requirements showing a need for at least 900 qubits to reach some speedup quantum advantage. p corresponds to the number of times the QAOA circuit blocks are repeated in the algorithm ansatz. It means that p=8 has a depth that is twice as large as for p=4. This would require physical qubit fidelities in the 99.9986% range, which is far out of scope for NISQ architectures. Source: G. G. Guerreschi and Anne Y. Matsuura [115].

Another talked about paper by Bao Yan et al (China) and related to the implementation of a QAOA-based algorithm to factor large integers didn't use the same precautions. It is based on using a classical "Schnorr" algorithm paired with some QAOA quantum procedure [128]. However, the paper doesn't provide any indication on the solution speedup and computing time estimations. It could be in the million years range for factorizing an RSA-2048 key. Also, they state that their QAOA algorithm would require only 372 NISQ physical qubits, giving the false impression that IBM's recently announced Osprey QPU with 433 qubits would fit the bill. Unfortunately, the QAOA algorithm used in that case has a 1139 to 1490 gates depth which would require physical two-qubit gate fidelities of 99.99982% or a $1.8 \times 10^{-6}$ error rate. As announced in May 2023, IBM Osprey's two-qubit gate fidelities are far off this level, below 98% [129]!

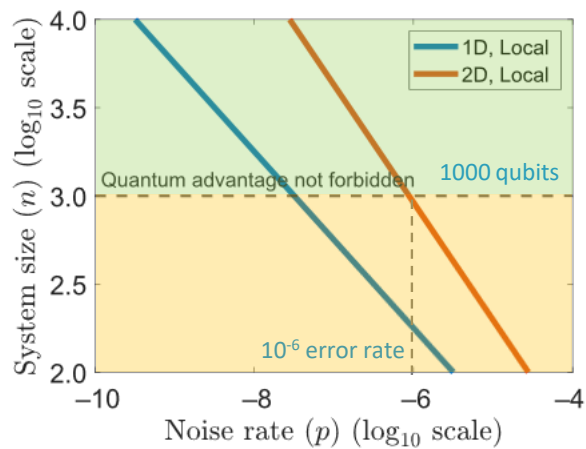

Figure 13: figure showing that a quantum advantage with QAOA would require two-qubit gate error rates in the $10^{-6}$ range, far from what is currently accessible. Source: Guillermo González-García et al [127].

Again, this would mean implementing some FTQC architecture with at least hundreds of thousands of physical qubits. The reaction from quantum information specialists like Scott Aaronson was abrupt, summarized in a "*No. Just no!*", more for theoretical reasons than for the practical ones mentioned above on real hardware resource needs[130].

### QML algorithms resources

In the literature, the situation seems not much better with quantum machine learning. The related algorithms running on NISQ are plagued with about the same problems than QAOA algorithms with regards to the way they could practically scale[131].

In November 2022, Lucas Slattery et al estimated that there is "*no quantum advantage with NISQ on QML with classical data*". Even worse, "*the geometric difference between "well-behaved" quantum models and classical ones is small and goes down with the number of qubits*"[132].

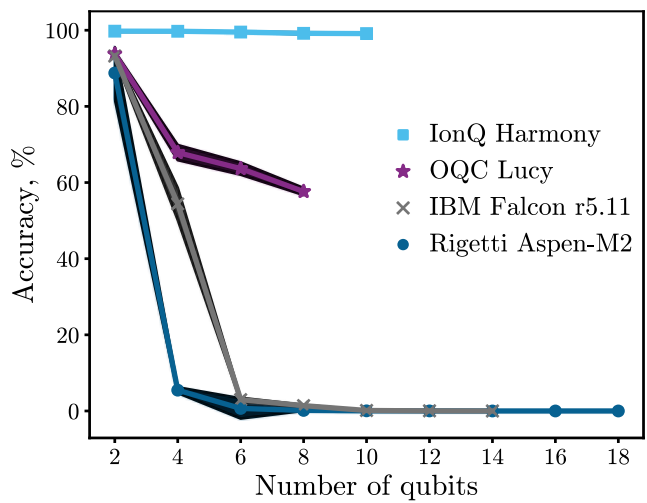

Figure 14: NISQ's actual algorithm depth with some current QPUs available with IBM and AWS cloud services when running some hybrid quantum neural network algorithm inferences (HQNN, proposed by Terra Quantum). It shows that the accuracy of the neural network predictions is trending to zero after 8 qubits for superconducting qubit platforms and is very good but capped at 20 qubits with trapped ion systems. We have here the illustration of the difficulty to have both breadth (number of qubits) and depth (number of gate cycles linked to qubit fidelities) in current NISQ platforms. Source: Kordzanganeh et al [17].

On existing QPUs, Mohammad Kordzanganeh et al found that the precision of a shallow quantum neural network training algorithm is below 10% when run with over 8 qubits for IBM and Rigetti as shown in Figure 14 [17]. It is better with OQC and IonQ but limited in number of qubits since these don't scale yet beyond 20 qubits and 20 qubits are cheaper to emulate classically whatever the scenario. So, we are very far from any quantum advantage, let alone doing something that cannot run on a simple laptop.

Other advances in QML algorithms are tested on QPUs with a very low number of qubits, like in the work of Diego H. Useche et al which "*presents a novel classical-quantum density estimation strategy for current noisy quantum computers, which combines quantum algorithms to compute the expectation values of density matrices with a new quantum variational representation of data called quantum adaptive Fourier features (QAFF)*". It was tested on an IBM Oslo QPU with 7 qubits and the discussion about its scalability seems absent in regards of these systems qubit gate fidelities[133].

Thanks to quantum algorithms dequantization, Jordan Cotler et al show "*that classical algorithms with sample and query (SQ) access can sometimes be exponentially more powerful than quantum algorithms with quantum state inputs*"[134]. For them, the only QML advantage can be obtained when the QPU has direct access to quantum data as input. Quantum algorithm dequantization consists in converting a quantum algorithm into a classical algorithm with decomposing it into subsets of tensor matrix operations that can be executed efficiently on a classical computer.

The purpose of dequantization is to run a given quantum algorithm more efficiently on a classical computer. Pioneering work in dequantization work was done by Ewin Tang in her thesis supervised by Scott Aaronson when she dequantized a recommendation algorithm under certain conditions in 2018[135].

Pradeep Niroula et al created a deep learning algorithm enabling the creation of documents summaries[136]. This hybrid algorithm had a classical part doing a lot of classical data preparation. It analyzed a dataset of 300,000 news articles from CNN and the Daily Mail and precomputed it with a BERT NLP (natural language processing) classical deep learning model that handled sentences extraction and their conversion into vectors. The quantum part managed the text summarizing from respectively 20 to 8 and 14 to 8 sentences, with Quantinuum QPUs H1-1 and H1-2 QPUs using respectively 20 and 14 qubits, and with a 100 qubit gates depth which is excellent. But we are not yet in the quantum advantage regime with this number of qubits which, again, can be emulated on a simple laptop, and probably faster on a server cluster! The paper doesn't provide resources requirements estimates for a larger summary set for, say, 100 or 1000 sentences. The way it was communicated was slightly exaggerated as shown in Figure 15.

## Constrained Quantum Optimization for Extractive Summarization on a Trapped-ion Quantum Computer

Pradeep Niroula[1,2,3,*], Ruslan Shaydulin[1,+,*], Romina Yalovetzky[1,+], Pierre Minssen[1], Dylan Herman[1], Shaohan Hu[1], and Marco Pistoia[1]

[1]JPMorgan Chase, New York, NY, USA
[2]Joint Center for Quantum Information and Computer Science, NIST/University of Maryland, College Park, MD, USA
[3]Joint Quantum Institute, University of Maryland, College Park, MD, USA
*These authors contributed equally.
+ruslan.shaydulin@jpmchase.com

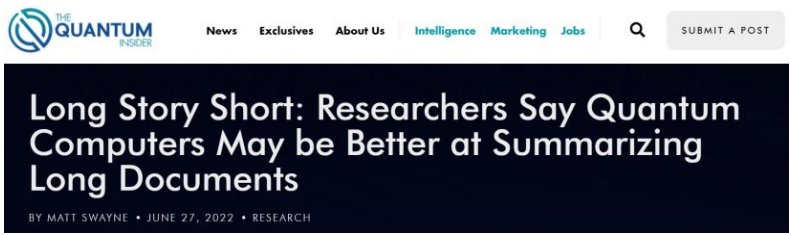

Figure 15: one exaggeration (below) about a QML algorithm's capacity to summarize long documents (source above). Sources: Niroula et al and The Quantum Insider [136].

In another work, Robin Lorenz, Bob Coecke et al implemented some natural language processing algorithm using over 100 sentences as entry with using only 5 qubits on an IBM QPUs[137]. Likewise, Wei Xia et al presented in March 2023 an improved quantum reservoir computing algorithm that could run on up to 7 qubits with some precision improvement over classical reservoir computing methods for forex forecasts.

But these 7 qubits run faster on a classical software emulator than on any existing QPU and the paper doesn't mention any qubit fidelities and number requirements to reach some quantum advantage[138].

Late 2022, Ismail Yunus Akhalwaya et al touted that NISQ systems would soon be able to solve topological data analysis problems (see Figure 16)[139]. TDA is used for extracting complex and shape-related summaries of high-dimensional data. NISQ-TDA was presented as the "*first fully implemented end to-end quantum machine learning algorithm needing only a linear circuit-depth, that is applicable to non-handcrafted high-dimensional classical data, with potential speedup under stringent conditions*". Practically speaking, TDA can identify clusters in high-dimensional data. It serves to estimate a "Betti number" which measures the connectivity of a topological space.

But we are far from being able to implement this algorithm in a NISQ regime[140]. It is a narrow implementation of the TDA class of problems, and it imposes stringent data conditions to generate any computing advantage. On top of this, a NISQ computing advantage would require overs 96 qubits with 99.99% two-qubit gate fidelities which are not in the radar yet as shown in Figure 17. Again, with even a fidelity of 99.9%, we would need at least about 9,600 such physical qubits.

Alexander Schmidhuber and Seth Lloyd "*argue that quantum algorithms for TDA run in exponential time for almost all inputs by showing that (under widely believed complexity theoretic conjectures) the central problem of TDA - estimating Betti numbers - is intractable even for quantum computers* [...] *Our results imply that quantum algorithms for TDA offer only a polynomial advantage*"[141], which, if implementable in a real NISQ regime would make sense. But given the overhead of FTQC that would be mandated to solve this class of problem, we'd have to look at the constants and other fixed costs to check that a quantum advantage would show up in a reasonable regime.

## Towards Quantum Advantage on Noisy Quantum Computers

Ismail Yunus Akhalwaya[1,5*]     Shashanka Ubaru[2,4*]     Kenneth L. Clarkson[3]
Mark S. Squillante[2]     Vishnu Jejjala[6]     Yang-Hui He[7]     Kugendran Naidoo[6]
Vasileios Kalantzis[2]     Lior Horesh[2]

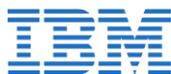

1. IBM Research, South Africa, Johannesburg, South Africa
2. IBM Research, T.J. Watson Research Center, Yorktown Heights, NY, USA
3. IBM Research, Almaden, San Jose, CA, USA
4. Oden Institute for Computational Engineering, and Sciences, University of Texas, Austin, TX, USA
5. School of Computer Science and Applied Mathematics, University of the Witwatersrand, Johannesburg, South Africa
6. Mandelstam Institute for Theoretical Physics, NITheCS, CoE-MaSS, and School of Physics, University of the Witwatersrand, Johannesburg, South Africa
7. London Institute for Mathematical Sciences, Royal Institution, UK
* These authors contributed equally to this work. Corresponding authors.

a large classical machine with 2 GPUs). The surface plot extrapolations provide the minimum noise-level requirements for NISQ-TDA to successfully run on future larger NISQ devices. The exciting prediction is that a 96-qubit quantum computer with a two-qubit gate and measurement fidelity of $\approx 99.99\%$ suffices to to achieve quantum advantage on the Betti number estimation problem.

Figure 16: topological data analysis quantum algorithms could reach a quantum advantage in a NISQ regime. Well, with 96 qubits with 99.99% two-qubit gate fidelities. Source: Akhalwaya et al [139]. Beware of any scientific paper with a title starting with "towards"!

Another older paper is more optimistic on TDA resource requirements. It states that a quantum TDA algorithm can have a guaranteed superpolynomial quantum speedup vs classical computing[142]. It says a quantum advantage would require at least 80 physical qubits but gives no precise indication on the algorithm depth. With the shallowest algorithm possible of 8 gate cycles, we still would need two-qubit gate fidelities in the 99.8% range.

On the other hand, quantum machine learning speedups are not the sole potential quantum advantage attribute but, as Maria Schuld and Nathan Killoran pinpoint, the comparisons are complicated between classical and quantum machine learning algorithms[143]. It deals with classifications quality, generalization capability on unseen training data, training data requirements and the likes, with few benchmarking references. On top of this, training data ingestion is mostly done by the classical part to prepare the algorithm quantum ansatz, and it scales linearly with the data size, so with no foreseeable quantum advantage.

At last, like VQE algorithms, QML algorithms have to fight the famous barren plateau problem, which prevents training convergence unless the ansatz circuit is really shallow[144]. It is the equivalent of avoiding local minima traps in classical machine learning, when a global minimum is searched but difficult to reach[145]. Research is very active to fix this problem like with adding additional parameters and constraints to improve gradients in the variational training loop without resorting to inefficient overfitting[146]. It also seems that the barren plateau syndrome can be avoided in VQE algorithms[147] [148].

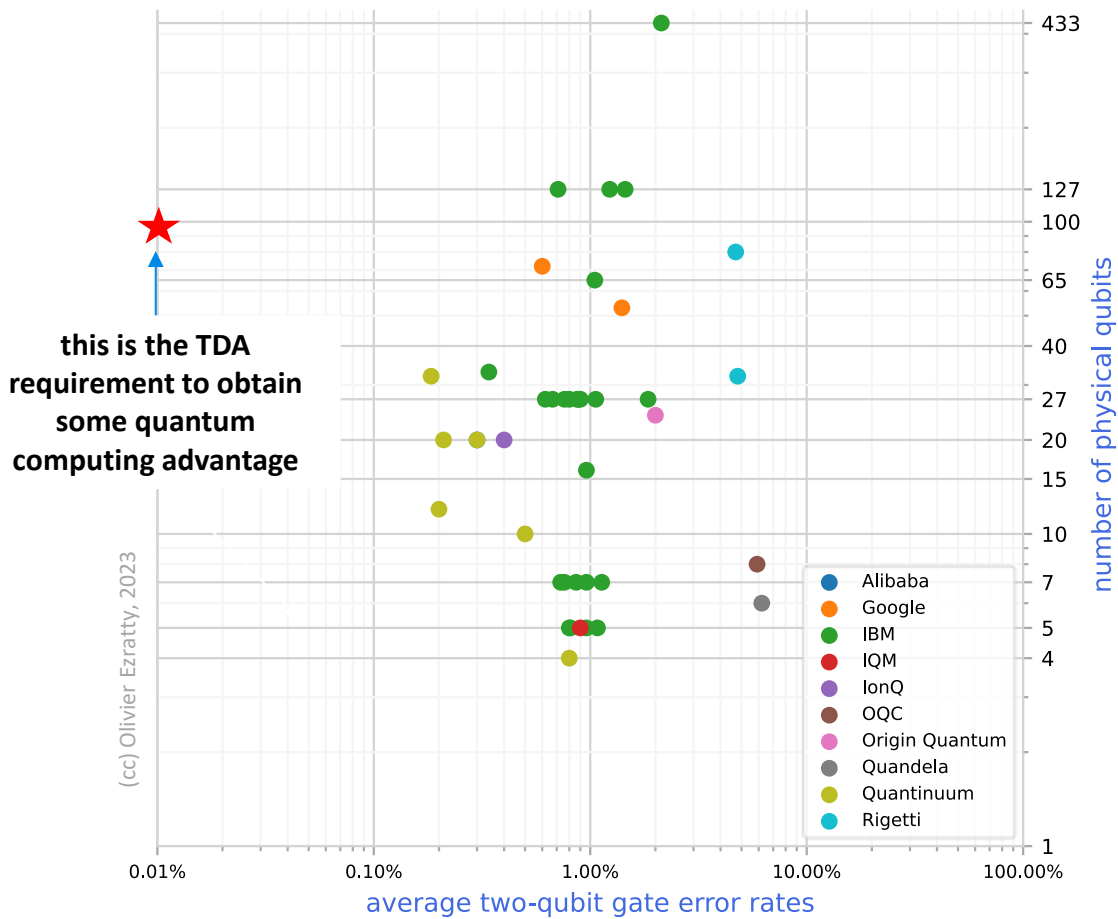

Figure 17: a topological data analysis algorithm (extracting complex and shape-related summaries of high-dimensional data) requirement of two-qubit gate fidelities of 99.99% is way outside currrent hardware capabilities. And this algorithm implementation is constrained by many specific requirements related to data sparsity. Source: Olivier Ezratty and vendors data as of May 12th, 2023.

## IV. NISQ POTENTIAL ENABLERS

So far, we've painted a rather gloomy picture of the whereabouts of NISQ, at least in the short term. Here, we'll discuss potential solutions even though they are still quite sketchy and unproven. How can some of the current weaknesses of NISQ QPUs be addressed so that they enable some form of quantum computing advantage?

We can frame these as about improving:

- **Quantum error suppression and mitigation techniques** although it is known that these techniques have an exponential cost with the circuit depth or qubit number (NISQ specific).

- **Algorithms resiliency** to noise and other hardware requirements constraints. This resiliency is quite rare and show up mostly with some particular quantum machine learning techniques (NISQ specific).

- **Scaling analog quantum computing** platforms given they have their own limits and belong to a side category in the NISQ realm (NISQ specific).

- **Qubit fidelities and capabilities** to enable larger quantum volumes and a larger number of high-fidelities qubits in the QPUs (not NISQ specific).

- **Qubit connectivity** to enable shallower algorithms implementations and faster computing times (not NISQ specific).

- **Quantum advantages** other than speedups (not NISQ specific).

- **Energetics** which could come out as being a key operational advantage of NISQ systems provided useful calculations are done in the first place.

We propose here a high-level assessment of these various techniques, some of these not being specific to NISQ architectures as highlighted above.

### Qubit fidelities and capabilities

Improving qubit fidelities is of course easier asked than done. All quantum computing research labs and industry vendors are working hard in that direction, with various results. As described in another paper related to Moore's law potential application to quantum computing, I looked at the various ways to improve this critical figure of merit[149]. It deals mostly with qubit design, primary gate set design, materials selection, manufacturing quality and control electronics signals purity. This quest is applicable to both NISQ and FTQC platform developments.

Qubit fidelities cover qubit initialization fidelity, single and two-qubit gate fidelities and qubit readout fidelities[150]. We'll focus here on two-qubit gates fidelities, that are also shown in Figure 5 and Figure 17.

The interesting QPUs, existing or prospective, are those which exhibit two-qubit gate fidelities that are over 99.5%. There are only a few at this point with a couple trapped ions and superconducting QPUs from IonQ, Quantinuum and IBM.

Trapped ions seem to have a hard time to practically scale to over 40 qubits. No single platform has so far reached 99.9% two-qubit gates fidelities, as well as for qubit preparation and readout. IBM has a goal of releasing its Heron 133 qubits processor with such fidelities in 2023.

Some alternatives are in the making:

- **Carbon nanotubes** spin qubits from C12 Quantum Electronics could reach the 99.9% threshold and have so far been digitally simulated.

- **Nitrogen and silicon carbide vacancy centers** are also good candidates for high-fidelity qubits although they are currently hard to manufacture at scale.

- **Photon qubits** have different figures of merit since they don't decohere natively. The trouble to fix is about their statistics and the need to have deterministic sources of photons, preferably assembled in cluster states of entangled photons, and with using deterministic photon detectors. These qubits can also significantly expand the computational space with multimode photons based on Fock numbers or frequency encoding. With Fock number photonic encoding, the Hilbert space can reach a size of $4^N$ instead of $2^N$ for N photons[151]. Also, some recent experiments in China made it possible to program a Gaussian Boson Sampler and solve some graph problem in a scalable way[152]. It still needs to be fact-checked thoroughly.

- The class of autonomously corrected qubits in the bosonic qubit family are also promising. Among these are the **cat-qubits** developed by Alice&Bob and AWS and other bosonic codes qubits developed by Nord Quantique and QCI. They have natively very low bit-flip error rates but high phase-error rates that require some error correction, landing these qubits directly in the FTQC realm[153]. But some researchers are proposing to use these qubits without error correction, like with QAOA algorithms[154].

- Likewise, **Majorana fermions** (or MZM, Majorana Zero Modes) qubits provide some form of self-correction but will be implemented only with fault-tolerant error correction schemes, when it works. They do not belong to the NISQ QPU class.

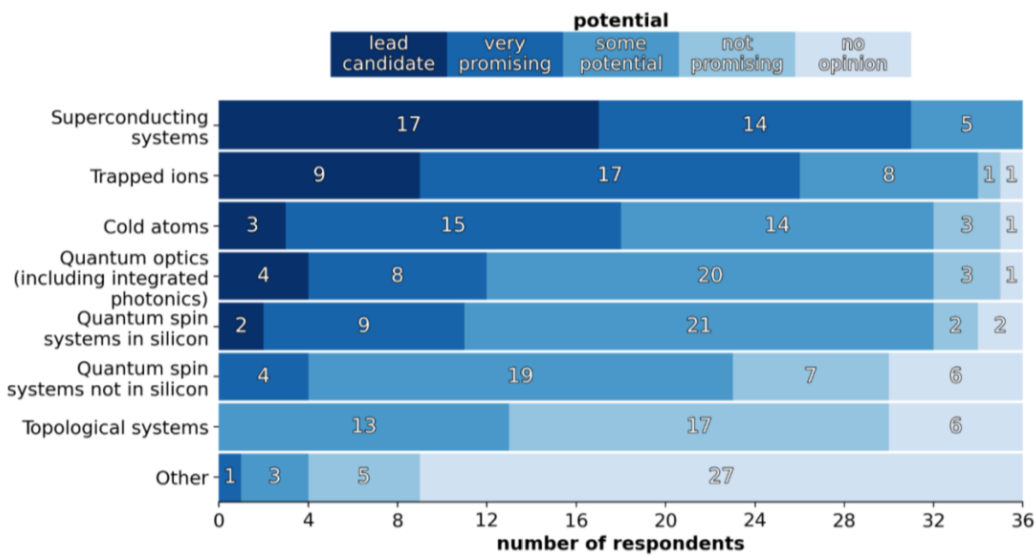

In the already mentioned 2022 Quantum Threat Timeline Report from the Global Risk Institute[21], quantum computing expert were polled on their opinion regarding the potential of physical implementation for quantum computer per type of qubits and their ability to support 100 logical qubits in the next 15 years, while it is not related to the viability of NISQ platforms, it provides a good indication of the platforms currently driving this kind of trust. It matches our inventory, although I wouldn't position cold atoms in the space of large scale gate-based quantum computing (see Figure 18).

### Qubits connectivity

Qubits connectivity plays a key role in minimizing the depth of many algorithms whether in NISQ or FTQC regimes, limiting for example the number of required SWAP gates in many algorithm implementations.

The best qubits with regards to connectivity are trapped ions. They showcase a many-to-many connectivity that on top of excellent fidelities make them a leading quantum computing platform.

It explains why trapped ions QPUs have the best quantum volume so far, in the $2^{20}$ range. Unfortunately, at this point in their development, these qubits don't scale well in number. All their current vendors (IonQ, Quantinuum, AQT, Universal Quantum, eleQtron) QPUs have under 30 qubits, and it is progressing very slowly.

Superconducting qubits have various types of connectivity as shown in Figure 19. The best ones are from D-Wave, although in quantum annealing mode, with clusters of qubits connected to 15 neighbors and soon 20 neighbors. Then, Google's Sycamore qubits are connected to 4 neighbors thanks to using tunable couplers. And finally, IBM's heavy-hex lattice enables a limited 1-to-2 and 1-to-3 connectivity.

Some quantum error correction codes like LDPC require long-range connectivity between qubits and it seems possible to implement it with stacked connectivity chipsets beneath the qubit chipset. Some progress could be expected here with adding more metal layers in the connectivity chipsets placed underneath the qubit's chipset. IBM and the MIT Lincoln lab are working on 3 and 7-layer connectivity chipsets to improve this connectivity for superconducting qubits.

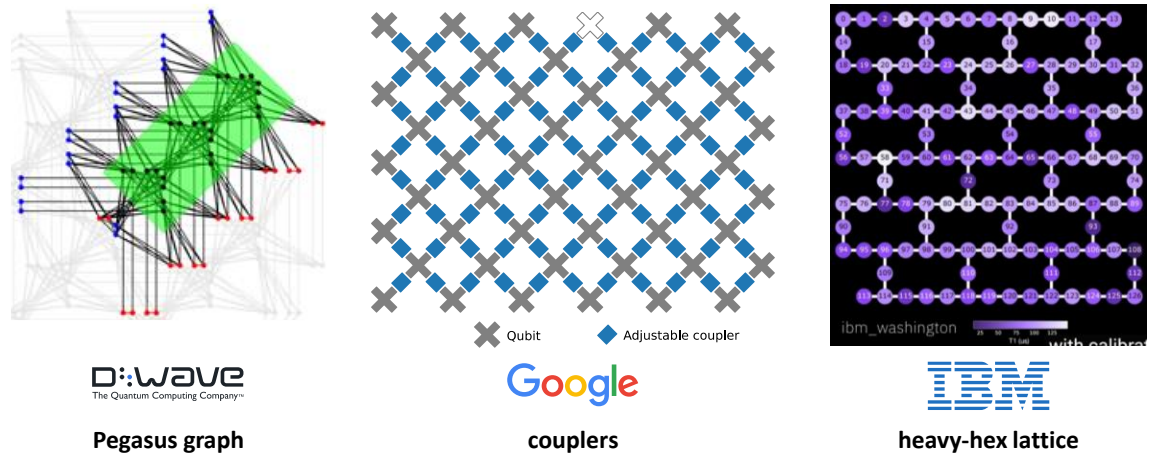

**Pegasus graph**        **couplers**        **heavy-hex lattice**

Figure 19: superconducting qubit connectivity difference across platforms. Sources: D-Wave, Google and IBM, even though D-Wave's connectivity in annealing mode can't really be compared with gate-based superconducting qubit circuits.

### Quantum error suppression and mitigation

Error handling is managed in different manners with quantum computers. The techniques used with NISQ systems are quantum error suppression and quantum error mitigation[155]. Fault-tolerant quantum computers will use quantum error corrections techniques which are not relevant for NISQ QPUs.

**Quantum error suppression** techniques deal with improving qubits at the physics level to minimize decoherence (loss of superposition and entanglement), cross talk (when actions on some qubits disturb other qubits) and leakage (when a qubit exits its $|0\rangle$ and $|1\rangle$ computational basis, which happens for example with superconducting qubits), and to maximize gate fidelities and speed. It also deals with qubit initialization errors and readout corrections. It is implemented with optimal electronics control (pulse shaping, reducing phase, amplitude and frequency jitter) and advanced device characterization and calibration. It depends on the type of qubit. When properly implemented, error suppression techniques scale relatively well with the number of qubits and algorithm complexity[156,157]. Error suppression techniques can also be used in FTQC settings. A variation error suppression technique is error filtration (EF) which is reusing a technique initially designed for quantum communications[158].

**Quantum Error Mitigation** (QEM) is about reducing quantum algorithms errors with combining classical post-processing techniques with some potential circuits modifications on top of running the algorithm several times and averaging its results (*aka* the "*expectation values of an observable*"). QEM reduces the influence of quantum errors using multiple runs and subsequent measurements coupled to some classical processing as opposed to QEC-based active qubits measurement and fast feedback-based corrections impacting the results of individual runs[159].

QEM proposals started to pop-up around 2016[160]. Most of them consist in learning the effects of noise on qubit evolutions and creating predictive noise models that can be applied to tune the results of quantum computations. Most QEM methods do not increase the required qubits count for a given algorithm.

Here are some identified QEM techniques:

**Zero noise extrapolation** (ZNE) builds error models based on solving linear equations. It supposes the noise is stable. It cancels noise perturbations by an application of Richardson's deferred approach[161] to the limit and works on short-depth (or shallow) circuits[162].

**Probabilistic error cancellation** (PEC) is about detecting circuit bias with finding noise quantum channels, represented as density matrices for quantum gates, using quasi-probability decomposition. There is a sampling overhead in the process. It then inverts a well-characterized noise channel to produce noise-free estimates of the algorithm observables (the 0s and 1s they're supposed to generate). It's also called Bayesian error mitigation and Bayesian read-out error mitigation (BREM).

**Learning Based Methods** QEM are based on machine learning techniques using training data to learn the effect of quantum noise in various situations[163].

**Error suppression by derangement** (ESD) which provides an exponential error suppression by increasing the qubit count by n≥2 but is still adapted to NISQ architecture and shallow circuits[157]. As with PEC, the regime where this method is useful is with very high fidelity qubits.

**Dynamical Decoupling** involves decoupling idle qubits from other qubits under certain conditions. It takes advantage of low level pulse control with superconducting qubits. It seems that under certain circumstances, it can generate a good quantum speedup for oracle-based algorithms. It has been tested on IBM and Rigetti QPUs[164 165 166].

**Other methods** include symmetry constraints verification, distillation using randomized benchmarking[167], randomized compiling[168], applying gates simulating the reverse effect of errors[169], depolarizing noise[170], quantum verification and post-selection[171], virtual distillation with derangement operators[172], using matrix product operators (tensor networks)[173], and mixing various QEM and QEC techniques[174].

There are also read-out noise mitigation and qubit readout error corrections techniques also known as measurement error mitigation techniques which are important for all variational algorithms[175 176].

Most of these QEM techniques have various limitations, including problems with accuracy and scaling[177], having computing time exponential overhead which in turn limits the potential quantum advantage of NISQ algorithms in its upper end regime[178]. However, these shortcomings may be limited in the narrow quantum advantage regime that NISQ could enable[179]. Viable NISQ may then show up in a narrow zone after a sufficient qubit fidelity is obtained above 99.9% and below the bad scaling effects of quantum error mitigation, as shown in the dotted ellipse in Figure 5 in page 8. This has yet to be determined theoretically and experimentally and separately for VQE, QAOA and QML algorithms variants.

## Algorithms advances

We've seen in the previous part on NISQ algorithms that their requirements are quite demanding to generate some quantum advantage. Most of them have been tested at a very low scale and would require a much larger number of qubits of much higher gate fidelities than are currently available and even foreseeable in the near to mid-term future.

Still, the improvements of algorithms design are encouraging. Many of them reduce the number of qubits and the gate depth requirements of typical variational algorithms (VQE, QAOA).

One example is an alternative to VQE that uses fewer qubits. This variational quantum selected configuration-interaction (VQ-SCI), is representing the target ground state as a superposition of "Slater determinant configurations" describing the wave function of electrons in molecules, encoded on the quantum computational basis states and making a preselection of the most dominant configurations. The algorithm has been tested with some of the usual suspect small molecules such as LiH, BeHe$_2$, NH$_3$, and C$_2$H$_4$ with up to 12 qubits. The number of qubits is equivalent to the number of spin-orbitals in the molecule[180]. However, as in many papers of this kind, there are no extrapolations on the qubit resource needs for larger molecules both in quantity and quality.

Another way to optimize molecular simulations with VQE is proposed by Algorithmiq and Trinity College (Dublin), using the ADAPT-VQE-SCF approach which combines a "self-consistent field approach" within the "Adaptive Derivative-Assembled Problem-Tailored Ansatz Variational Quantum Eigensolver (ADAPT-VQE) framework"[181 182]. They expected these techniques to yield useful quantum advantages in 2023. ADAPT-VQE can also be used to simulate atom nuclear shell models[183].

Always in the VQE domain, a team of German and Spanish researchers found a way to improve an algorithm handling flight gates aircraft assignment (FGA) in airports[184]. This algorithm goal is to minimize "*the total transit time of passengers in an airport by finding an optimal gate assignment of the flights*".

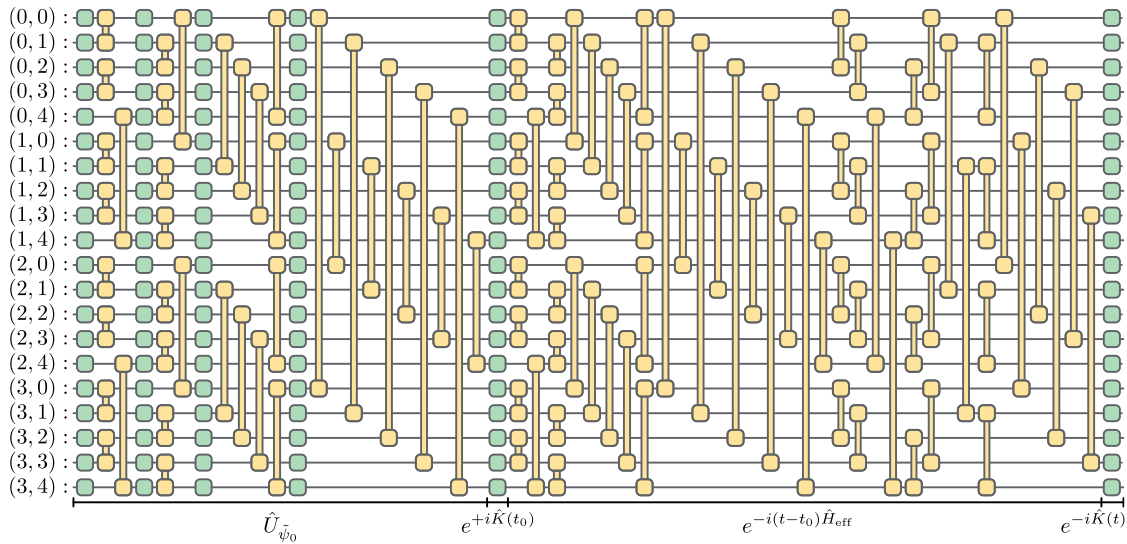

It requires fewer qubits in a NISQ architecture thanks to using a Conditional Value at Risk (CVaR) aggregation (classical) cost function that avoids a dominant subspace of invalid solutions. The ansatz used in the algorithm is shallow and quite classical. The algorithm improvement sits in its classical part. This CVaR-VQE algorithm has been tested with only 18 qubits so far, but with simulations showing a non-exponential growth in consumed resources as the problem size grows. Still, the related paper doesn't provide any indications on the real-life problem sizing and the classical and quantum computing resources needed to solve it properly. Regular non-variational algorithms also exist for some physics simulations. A recent one from Timo Exckstein et al is simulating dynamic quantum physics and was tested with a 40-gate cycle depth over 20 trapped ion qubits, as shown in Figure 20[185]. It's a record in its class but we are still dependent on scale to obtain some sort of quantum advantage. With 20 qubits, we are still able to run it faster on a simple laptop.

There is also a variational equivalent of the FTQC HHL (linear algebra equation) algorithm to solve the Quantum Linear Systems Problem (QLSP). The inverted matrix must be sparse. An increased precision is obtained but with only 4 qubits without some indication that it would scale well with the number of qubits[186].

Some progress is also made in improving the efficiency of NISQ QAOA algorithms although their resource estimations are frequently missing in the literature. Testing it on 10 qubits of a 27 qubit IBM QPU is insufficient[187].

At last, one relatively exotic way to obtain a quantum advantage with NISQ is to directly feed the QPU with quantum data, which can be done with using quantum sensors in theory, as shown in Figure 21[188]. It was implemented in 2021 with 40 superconducting qubits and 1,300 quantum gates running a QML algorithm. It is interesting but reserved for very specific use cases.

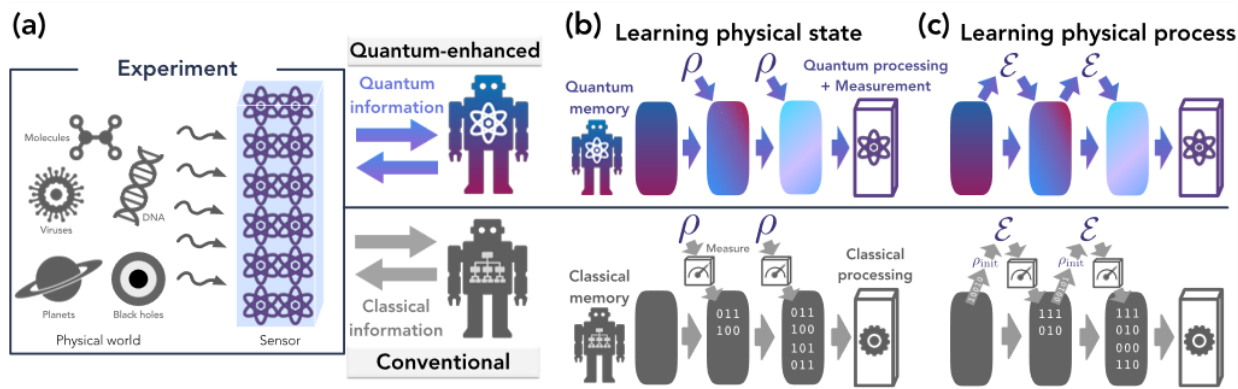

## Scaling analog quantum computers

Quantum annealing and analog quantum computing are not darlings of the quantum computing industry. On one hand, in quantum annealing, D-Wave has been criticized for a long time for "not being quantum" nor being in position to bring any computing advantage. On the other hand, analog quantum computers (programmable Hamiltonian simulation or programmable quantum simulators) are developed and commercialized by a very small number of vendors such as PASQAL and QuEra and said to have their own scalability challenges.

Still, when you compare objectively the documented case studies around, you find many solutions that are not far from reaching some quantum advantage[189]. Most of them are not yet "production grade" but they are closer to this status than all NISQ based prototype algorithms.

On top of that, recent benchmarks show that analog quantum computers currently have greater computing capacity than gate-based noisy quantum computers. In a 2022 paper, Ward van der Schoot et al evaluated the Q-score of D-Wave in various situations[190]. The Q-score from Eviden (Atos) measures the maximum size of a standard optimization problem (Max-Cut) that can be solved on a given system. The paper authors found that a D-Wave 2000Q had a Q-score of 70 and the more recent D-Wave Advantage has a Q-score of 140 with a classical and quantum annealing computing time limit of 60 s. The Q-score benchmark is based on a QAOA optimization hybrid algorithm for gate-based systems and can be implemented with a QUBO algorithm on annealers and quantum simulators.

Then, they evaluated Q-scores for hybrid solutions using a "tabu search" and obtained Q-scores of 12,500, while a single PC server could reach 5,800 and the quantum annealer alone, 2,300. Meanwhile, current gate-based quantum computers don't have Q-score above 20.

Another benchmark found a Q-Score of 80 for a PASQAL analog quantum computer, although it was determined with using a classical emulator (or "digital quantum simulator") of the quantum processor[191]. PASQAL even announced in February 2023 EMU-TN, a tensor networks-based emulator to simulate its programmable Hamiltonian simulator up to 100 atoms, noise included, and to estimate the resources required to run a given algorithm[192 193]. This is quite encouraging for the prospects of quantum annealing and analog quantum computers. But how are we with regards to actual practical case-studies and how does the technology scale compare to gate-based NISQ and FTQC?

Sheir Yarkoni et al's review paper on quantum annealing provides an up-to-date status of the D-Wave platform usability[194]. It describes how optimization and graph problems are mapped onto the D-Wave QPU qubit structure, through the process of minor embedding. It inventories a broad set of algorithms and trials related to mobility traffic flow optimization and vehicle routing problem, scheduling and logistics problems, finance portfolio optimization, quantum simulation, chemistry and material design, physics, biology, machine learning (classification, reinforcement learning, cluster analysis), matrix factorization and other finite-element design. All these algorithms are hybrid like most NISQ known algorithms. Compared to the various known gate based NISQ algorithms, annealers are more generic than usually thought.

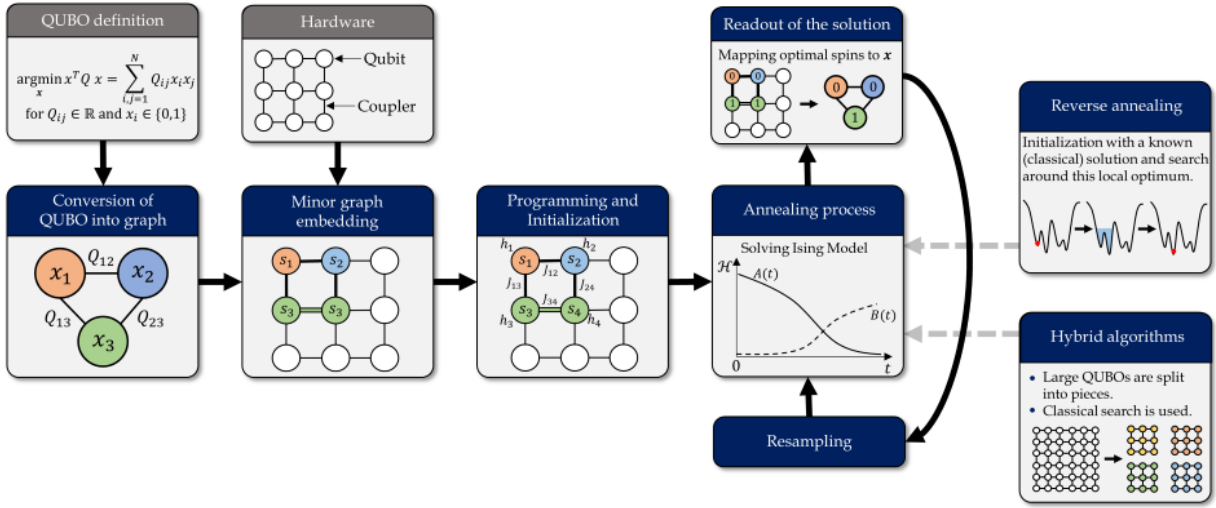

Figure 22: quantum annealing proccess taking the example of an optimization problem encoded in a QUBO problem and graph (quadratic unconstrained binary optimization). The graph is then automatically converted in a graph corresponding to the superconducting topology of a D-Wave QPU through the process of "minor graph embedding". The graph is then encoded in the system with an initialization of the qubit weights and connections. The annealing process takes place with setting a transverse magnetic field that sets the qubits value in a superposed state and progressively removing this field, which implements the annealing proccess, converging the qubit towards their optimum value minimizing the total system energy. Then the qubits are read out. The process is repeated several times and the result values averaged. Variations involved reverse annealing when the graph initialization uses a known classical solution and the annealing proccess helps find a better solution around it, and hybrid algorithms where large QUBO problems are split classically into smaller QUBO problems. Source: Sheir Yarkoni et al [194]

Yarkoni et al still highlight the various limitations of quantum annealing: it can only address specific problem formulations and it implements metaheuristic quantum optimizations, meaning approximate solutions (Figure 22). Also, the number of needed physical qubits scales polynomially with the number of logical variables from the problem formulation. Other sources found gate based QAOA to be potentially more efficient than quantum annealing based QUBO[195]. The review paper then lists some improvements required at the hardware level to generate some quantum advantage such as additional qubit control, driver Hamiltonians and operators, and higher connectivity. The next generation of D-Wave annealers is supposed to meet some of these requirements.

| Method | XS | S | M | L | XL | XXL |
|---|---|---|---|---|---|---|
| VQE | 2.4 % | - | - | - | - | - |
| Exhaustive | 5.1 % | 13.9 % | - | - | - | - |
| VQE Constrained | 5.1 % | 9.1 % | 7.1 % | - | - | - |
| Gekko | 5.8 % | 13.9 % | 13.6 % | 54.1 % | 71.6 % | - |
| D-Wave Hybrid | 5.8 % | 13.9 % | 13.6 % | 18.9 % | 29.3 % | 67.6 % |
| Tensor Networks | 5.8 % | 13.9 % | 15.4 % | 38.2 % | 39.6 % | 39.7 % |

TABLE III. Profits (percentual) computed by the different methods for the different datasets and time periods from Table I.

| Method | XS | S | M | L | XL | XXL |
|---|---|---|---|---|---|---|
| VQE | 278 | - | - | - | - | - |
| Exhaustive | 0.005 | 34 | - | - | - | - |
| VQE Constrained | 123 | 412 | 490 | - | - | - |
| Gekko | 24 | 27 | 21 | 221 | 261 | - |
| D-Wave Hybrid | 8 | 39 | 19 | 52 | 74 | 171 |
| Tensor Networks | 0.838 | 51 | 120 | 26649 | 82698 | 116833 |

TABLE IV. Run-times (in seconds) estimated for the different methods for the different datasets from Table I.

Figure 23: comparison of solution accuracy and runtimes for a dynamic portfolio optimization between gate-based VQE, D-Wave annealing and classical tensor networks, showing a potential quantum advantage on large problems profits and runtime with a D-Wave 2000Q processor. The hybrid solution seems to generate the best results while VQE solutions are limited by the number and quality of available gate-based qubits. However, the high portfolio profits generated here are subject to caution. Source: Samuel Mugel, Roman Orus et al [196].

In detail, many algorithms tested on quantum annealers are able to solve sizeable problems, but usually, still under the demanding quantitative requirements levels of real-life scenarios.

Let's take a couple examples coming mostly from the financial sector:

- Samuel Mugel et al compare implementations of portfolio optimizations with classical tensor networks, hybrid quantum annealing and a NISQ VQE algorithm running on IBM gate based QPUs[196]. They got the best results and largest calculations with the two first solutions, handling 55 assets over 8 years. See the related data in Figure 23.

- Samuel Mugel et al also created a variant for investment optimization with a minimal holding period constraint with handling 50 assets over a one year period, all using a D-Wave 2000Q[197]. It requires a few minutes of computing per day.

- Salvatore Certo et al from Deloitte handled a SP500 portfolio optimization with comparing CPLEX (classical optimization), BQM (a QUBO binary quadratic model) and CQM (QUBO Constrained Quadratic Model)[198]. See Figure 24 for performance comparisons.

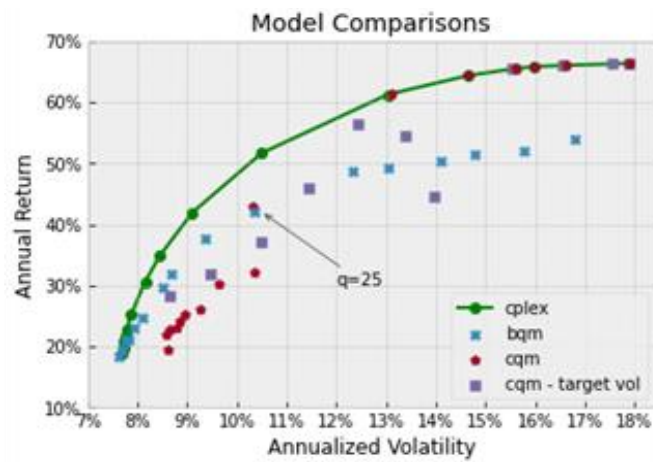

Figure 24: classical solutions (CPLEX, green dots, which is an exact solution) are performing better than various hybrid solver D-Wave solutions (blue, purple and red dots, providing approximate solutions) for optimizing a Standard & Poor's 500 stocks porfolio. Only in high volatility situations do quantum annealing generate solutions as good as classical ones. Source: Salvatore Certo et al [198].

- Samuel Palmer et al developed a financial index tracking algorithm with a 25 assets portfolio from the NASDAQ 100[199].

- Hristo N. Djidjev proposes two quantum annealing based methods to solve the set cover problem, which outperforms classical methods[200].

- Martin Vesely et al found out that a D-Wave annealer was close to solving various financial optimization problems like the determination of the optimal currency composition of foreign exchange reserves, while no single gate-based QPU was in such position[201].

Now onto analog quantum computers. They bring more flexibility on paper with the ability to define arbitrary graph trees with better connectivity.

- G. Semechin et al simulated the physics of some topological spin liquid using a 219-atom analog quantum computer from Mikhail Lukin's Harvard's lab (QuEra)[202].

- Lucas Leclerc et al from PASQAL, Multiverse and CACIB used a QBoost hybrid algorithm using a PASQAL neural atoms-based analog quantum computer to predict « fallen angels », which are the companies who could fail in loans reimbursements[203]. The data set used 20 years of historical data containing 90 000 items with 150 features on 2000 companies organized in 10 verticals and 100 sub-verticals from 70 countries. The training data set used 65 000 items while tests were done on 26 000 items. The study found out that a quantum advantage could show up with 150 to 342 neutral atoms when compared to a best-in-class classical machine learning models, and 2.800 atoms for the more precise subsampling method.

- Various analog quantum computing algorithms do not need variational loops like with VQE and QAOA on gate-based quantum computers. Examples were proposed with kernel based QML algorithms for solving graph problems[204], supervised learning and partial differential equations solving which evaluates once, but trains classically[205], and classical training of quantum sequences for which probability amplitudes at the output are known, yet sampling from the distributions that would be prepared by an actual quantum computer is classically hard to simulate[206].

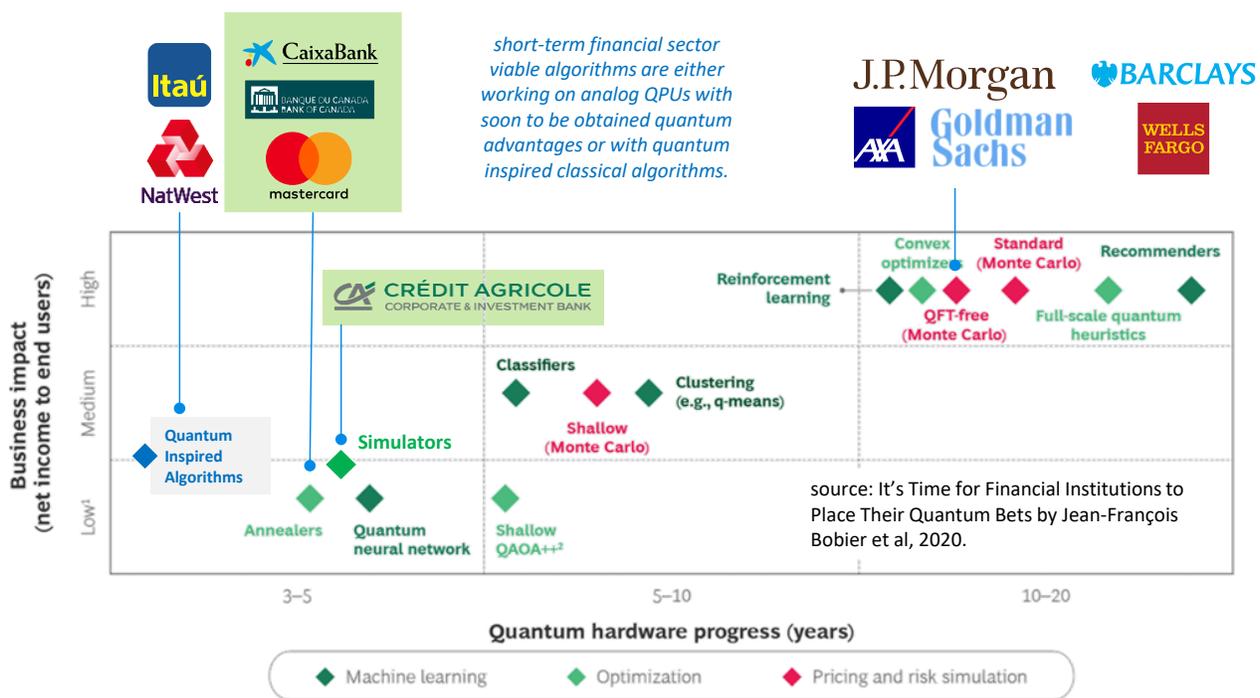

Figure 25: where are the case studies in the financial sector: with quantum annealing, quantum simulation and quantum inspired algorithms. As of 2023, short term case studies with small business impact can run with classical quantum inspired algorithms. Prototype solutions have been created with quantum annealers and quantum simulators but do generally not yet scale to production grade levels. At last, the most interesting business use cases and algorithms require fault-tolerant quantum computers with thousands of logical qubits and are positioned in the very long term accordingly, given these systems scalability is yet to be demonstrated practically. . Source: BCG chart[207] and Olivier Ezratty additions.

When reviewing all these case studies, both in the gate-based and analog quantum computing categories, one thing is striking: the most powerful solutions available are in the analog space rather than in the gate-based space. Quantum inspired classical solutions implementing linear algebra and tensor networks computing are also making classical computing more competitive in several areas (see Figure 25)[208]. These are not quantum at all.

Then, other use cases directly put you in the FTQC zone, requiring thousands of logical qubits and thus, millions to hundred million physical qubits.

However, even if analog quantum computing existing use cases are closer to real-life production grade levels than the gate-based equivalents, there are still some challenges to overcome in generating a quantum advantage with analog quantum computers as summarized in Figure 26.

Quantum annealers require more tunability of qubit connections and better qubits connectivity. Noise mitigation must also be handled [209] [210] [211].

And there's a remaining question similar as the one with NISQ systems: how far can large scale quantum effects work, particularly, based on the tunnel effect that is at the core of quantum annealing.

With neutral atoms, their scaling is linked to the ability to control large chunks of well-positioned entangled atoms in ultra-vacuum. The related tools are made of more powerful and stable lasers, and their related control electronics. Also, the research grade optics table used to control all the quantum computer device will have to be redesigned to avoid the tedious positioning tuning done to set up and calibrate these QPUs.

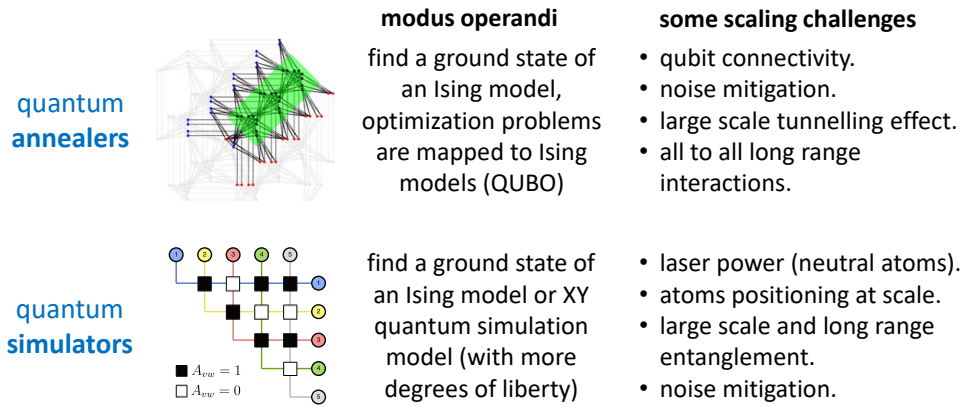

| | **modus operandi** | **some scaling challenges** |
|---|---|---|
| **quantum annealers** | find a ground state of an Ising model, optimization problems are mapped to Ising models (QUBO) | • qubit connectivity.<br>• noise mitigation.<br>• large scale tunnelling effect.<br>• all to all long range interactions. |
| **quantum simulators** | find a ground state of an Ising model or XY quantum simulation model (with more degrees of liberty) | • laser power (neutral atoms).<br>• atoms positioning at scale.<br>• large scale and long range entanglement.<br>• noise mitigation. |

Figure 26: some scaling challenges for quantum annealers and analog quantum computers. Source: (cc) Olivier Ezratty, 2023.

### Other NISQ techniques

Let's now make a small inventory of various techniques that could potentially make NISQ viable although their assessment is still in progress since, most of the time, they have not been practically experimented.

**DAQC** (Digital-Analog Quantum Computing) is a proposal to implement a hybrid gate-based and analog quantum computing model[212]. DAQC is supposed to make a more efficient use of quantum computing resources and enable NISQ algorithms with fewer qubits and to run faster than regular NISQ QPUs. It is adapted to optimization and machine learning. It is proposed by Kipu Quantum (Germany) and Qilimanjaro (Spain). Kipu Quantum is investigating the use of superconducting, trapped ion and neutral atoms qubits. QPU chipsets would have custom designs to handle global entangled states for the annealing part. An implementation proposal using superconducting qubits would use SQUIDs to connect qubits in 2D matrices[213]. It can improve computing fidelities to some extent[214]. Questions abound on the speedups obtained with this architecture, its dependance on algorithms classes and its impact on control electronics and energetics. Also, it is more complicated to debug algorithms and few development tools are supporting it. In a recent paper, Narendra N. Hegade and Enrique Solano could factorize a 48 bits integer on 10 Quantinuum qubits and asserted that a DAQC NISQ platform could enable a factorization of RSA-2048 keys[215].

**LHZ architecture** (for its inventor names: Lechner–Hauke–Zoller) developed by ParityQC (Austria) using sort of small logical qubits in a variation of quantum annealing that makes it programmable[216]. The architecture can be implemented using superconducting, NV-centers, quantum dots, and neutral atom qubits[217]. ParityQC proposes a related technique to reduce QAOA errors with quantum error mitigation[218].

**Circuit cutting** and **entanglement forging** are two NISQ techniques proposed by IBM Research.

Circuit cutting splits "*a quantum circuit into multiple smaller circuits with fewer qubits and gates such that the result of executing the collection of the smaller circuits is the same as the result of executing the original circuit by exploiting subsequent classical postprocessing*". It can be implemented to improve QAOA expectation values but the benefit decreases with the graph size[219]. Tests were done with IBM 27 qubit QPUs and not in a regime of potential quantum advantage. Also, aren't the smaller circuits easier to emulate classically, thus cancelling any quantum advantage? It is also proposed to optimize quantum simulations[220].

Entanglement forging "*harnesses classical resources to capture quantum correlations and double the size of the system that can be simulated on quantum hardware.*". It is mostly used with VQE for molecular simulations or quantum machine learning and is based on a Schmidt decomposition and SVD (singular value decomposition) of a quantum state into a bipartite state of N+N qubits[221]. It was tested on a 5-qubit QPU and its scalability has to be demonstrated. If a quantum system can be decomposed into two separable states, does it mean we are halving the size of our computing Hilbert spaces, thus losing a lot of quantum entanglement gains?

A third technique, circuit knitting is clustering the circuit into high-interaction parts on the same QPU, across a multi-core and distributed architecture with some quantum communication like microwave links or photon entanglement links. In theory, this technique proposed by IBM would enable a full Hilbert space with a size of $2^N$, N being the total number of qubits[222].

**Q-CTRL** (Australia) provides a quantum control infrastructure software working at the low-level firmware level controlling qubit drive microwave pulses, using machine learning to improve these qubits control pulses and optimize quantum error correction codes (see Figure 27). It is a quantum error suppression technique.

Their Python toolkit is used by quantum computers designers working with IBM Qiskit, Rigetti and with Quantum Machines microwave pulse generators. They implement error-correction techniques that increase the likelihood of quantum computing algorithm success between 1000x and 9000x on quantum hardware, as measured using the QED-C algorithmic benchmarks.

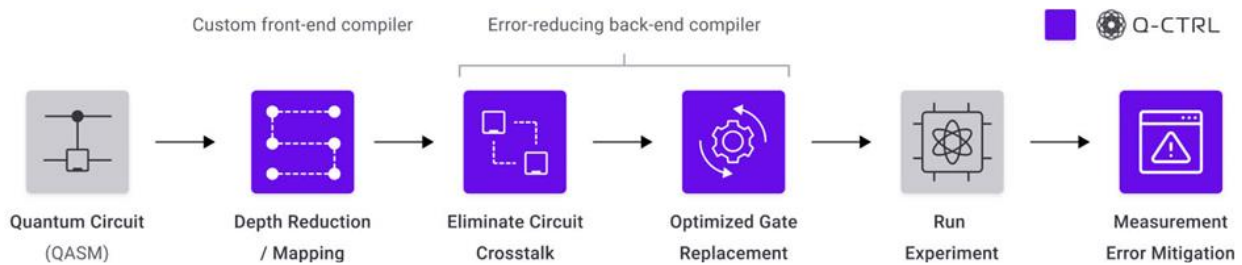

Figure 27: the Q-QTRL Boulder Opal architecture to optimize superconducting qubit control pulses. Source: Q-CTRL.

**NISQ+** is a technique proposed in 2020 by Intel, the University of Chicago, and the University of Southern California (USC), that is using fast approximate quantum error correction and quantum error mitigation, SFQ superconducting control electronic circuits running at 3.5K and lightweight logical qubits[223]. It is intermediate between NISQ and FTQC. It could augment the usability of NISQ QPUs by several orders of magnitude. It could for example extend the computing depth of 40 to 78 qubits QPUs to millions of gate cycles with using only 1000 physical qubits as shown in Figure 28.

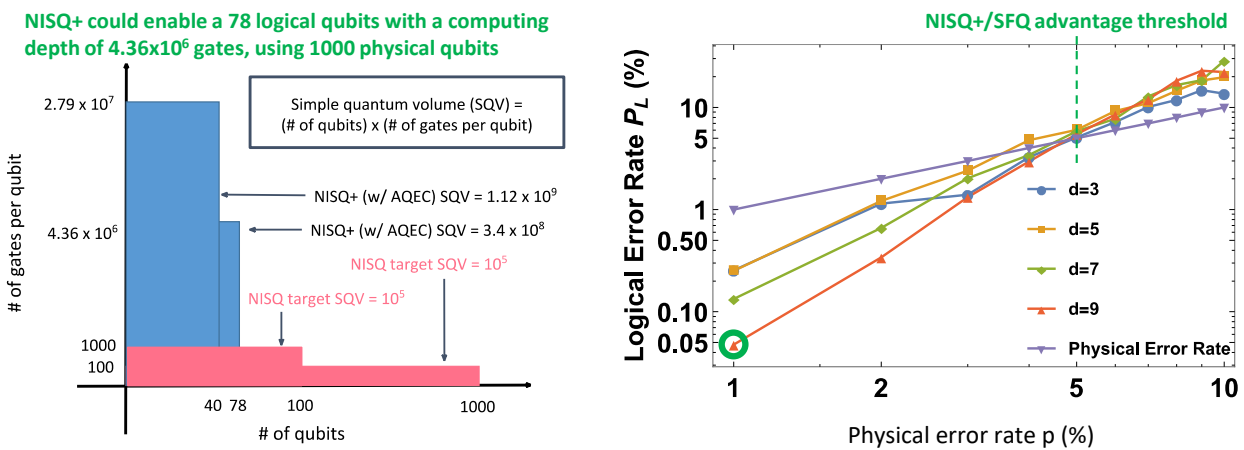

Figure 28: NISQ+ could potentially enable the creation of 78 logical qubits and a good computing depth. They use the notion of Simple Quantum Volume (SQV) which is the qubit number times their available gate depth. Still, satisfying logical qubit error rates require a surface code of distance 7 to 9. Source: Adam Holmes et al [223].

**Finding other quantum advantages**

Most quantum algorithms are created with the goal of achieving a quantum speedup over the best-in-class classical counterparts. This computing time speedup is usually theoretically either polynomial or exponential, the Holy Grail being an exponential speedup.

But practically, it seems that most NISQ algorithms have only at most a moderate polynomial speedup. And the cross-over with best-in-class classical algorithms may happen at very large time thresholds due to high constants in the quantum regime and to the rather slow gate cycles of quantum computers. Meaning, a NISQ implementation is better than the classical regime with computing times exceeding days, months and even years. The difference may even be minimal if the classical part of variational algorithms computing is very long and doesn't scale well[224].

| advantage | | definition proposal |
|---|---|---|
| space | → | when the qubit register data space - scaling in $2^N$ complex numbers with N qubits - exceeds the memory capacity of classical computers. |
| speed | → | when a fully-burdened quantum algorithm, including its classical part, runs faster than an equivalent best-in-class entirely classical algorithms running on either the largest supercomputers or a given HPC configuration. |
| quality | → | when the quality of the results of a quantum algorithm is better for some respect than the best-in-class classical algorithms. It can relate for example to the error rate of some machine learning classification or to the precision of a chemical simulation to find the ground state energy of a many-body system. |
| energetic | → | when a fully-burdened quantum computer and algorithm configuration consumes less energy than the best-in-class classical equivalent. It becomes a sort of energetic supremacy if no classical computing configuration can solve the given problem. |
| cost | → | when the total cost of the quantum solution is lower than the total cost of a best-in-class classical solution. There are many ways to calculate this cost. It can be just about hardware and software or also include other incurred costs like people training and cost of software development. |

Figure 29: a proposal set of definition for the various advantages when comparing quantum and classical quantum settings. When making any comparisons, quantum settings should include all their surrounding classical computing environments. Also, a comparison can be made with either the largest supercomputer in the world or with a smaller classical computing setting, like a mid-size HPC system. In the end, the business benefit will come from a given balance of cost-speed-quality benefits and trade-offs. Source: (cc) Olivier Ezratty, 2023.

However, in some situations, NISQ quantum algorithms could help create better solutions than their classical counterpart. But it is difficult to evaluate, particularly with QML.

Some qualitative aspects generated by NISQ solutions could be a better precision of predictions and classification for QML, less training data for QML, or better heuristic results for optimizations implemented in QAOA or physics simulations with variations of VQEs. The other potential advantage is the favorable energetics of quantum computers. But to assess it, any NISQ quantum algorithm must be able to do at least as well as the best-in-class classical algorithms, given the comparisons are never easy to make.

Comparing quantum and classical computing systems is more subtle than just looking at speedups. We propose in Figure 29 a taxonomy of these various types of quantum advantages, with precising that the classical comparison point may not necessarily be the largest supercomputer available. Also, this taxonomy is not bound to be theorical with looking at asymptotic polynomial or exponential advantages, but on practical advantage with given sets of algorithms and real use cases using production-grade input data sets.

Many papers discuss these aspects, but they fail to account for the real state of the art of classical computing. There is a lot of work to do in that space with more theoretical and experimental data, and better precision on the classical computing equivalences used in the comparisons. This is particularly true for QAOA combinatorial optimizations where the top notch classical algorithms are rarely accounted for.

We need to define some form of "quantum equivalence" when a quantum system is at least as good as its full classical counterpart but not necessarily superior. It's a fine threshold which also depends on the reference classical hardware, which is not necessarily in the top 100 worldwide supercomputers.

Questions abound here about cases, mostly in QML, where the ability to explore large Hilbert spaces helps create better quality results with NISQ quantum algorithms.

## NISQ energetics

If and when NISQ algorithms can show some superiority or even, just parity, in terms of speed over various classical settings, it will be interesting to compare their energetics. We may end up with a rather surprising outcome: one key benefit of NISQ platforms being a lower energetic cost when compared to their classical equivalent.

As we've seen so far, the only NISQ QPUs of interest are currently from IBM, D-Wave, PASQAL and QuEra. We must look at their roadmap in the NISQ era to see whether they could bring some computing and some energetic advantage altogether.

IBM and their 1,386 qubits Flamingo system to be released in 2024 could be interesting, while Condor's 1,121 qubits platform will probably not have sufficient fidelities to successfully run NISQ algorithms. With PASQAL and QuEra, we have to consider their next generations of neutral atoms analog quantum computers with 300 to 1000 actual controllable atoms. Other interesting QPUs to consider are multimode photon-based such as those from Quandela, and other systems from Xanadu[225].

The table in Figure 31 provides very rough indications of existing QPUs and future QPUs energetic footprints, with some additional details in Figure 30. To understand IBM's future Flamingo platform's estimate of 140 kW, we can guess that it will use a Bluefors KIDE cryostat containing 9 pulse tubes with Cryomech compressors, each consuming about 10 kW, plus a mutualized external water-water cooler for the compressors[226]. Then three dilutions will consume about 1 kW each in their gas handling system and control systems. Plus a few PCs, a vacuum pump and the control electronics consuming about 20W per qubit.

Power consumption cannot be directly compared with classical counterparts since computing time has to be considered. The energy footprint is not power, it is power × time. To estimate this footprint, we need to compute the number of gate cycles a given QPU would require for an algorithm and multiply it by the average gate length. This could yield an estimated power consumption in Joules per calculation. Work is of course needed to identify calculations of this type that perform as well as their equivalent classical counterparts, and then benchmark their respective power consumptions.

If we consider for example the case of IBM's future Flamingo platform with an estimated power drain of under 140 kW, it may compare favorably with HPCs if it can run NISQ algorithms successfully with reasonable ansatzes optimization cycles.

Then, you have three dilutions with a power of about 1 kW in their GHS (gas handling system) and control systems. Plus a couple PCs, a vacuum pump and the control electronics with a reasonable power budget of 20W per physical qubit. But all of this must be simulated, tested and computed before landing a conclusion. This is an open area of research and benchmarking.

Work is needed to identify algorithms performing at least as well as their equivalent classical counterparts. We could also find situations where an energetic advantage is significant, and a computing advantage is minor or non-existent. This could even show up in comparisons without any quantum computational advantage. A GPU based server cluster consuming up to 12 kW emulating about 40 qubits in vector state mode could consume much more than an equivalent NISQ quantum processor, like with photons or NV centers qubits, providing their noise doesn't make the comparison moot.

The comparison must consider the cost of quantum error mitigation and the impact of various compiler optimizer and transpilers improvements (transpilers convert code quantum gates into the primary quantum gates supported by the QPU). Finally, researchers and technology developers will need to identify potential full-stack power optimizations of their systems.

Finally, this makes sense by comparing these quantum systems with classical systems that work in similar functional regimes. We know for example that a full rack of Nvidia DGX has a power of about 30 kW and the largest supercomputer, the Aurora Frontier from the Oak Ridge National Laboratory from the US Department of Energy has a power of 22 MW at full scale utilization[227].

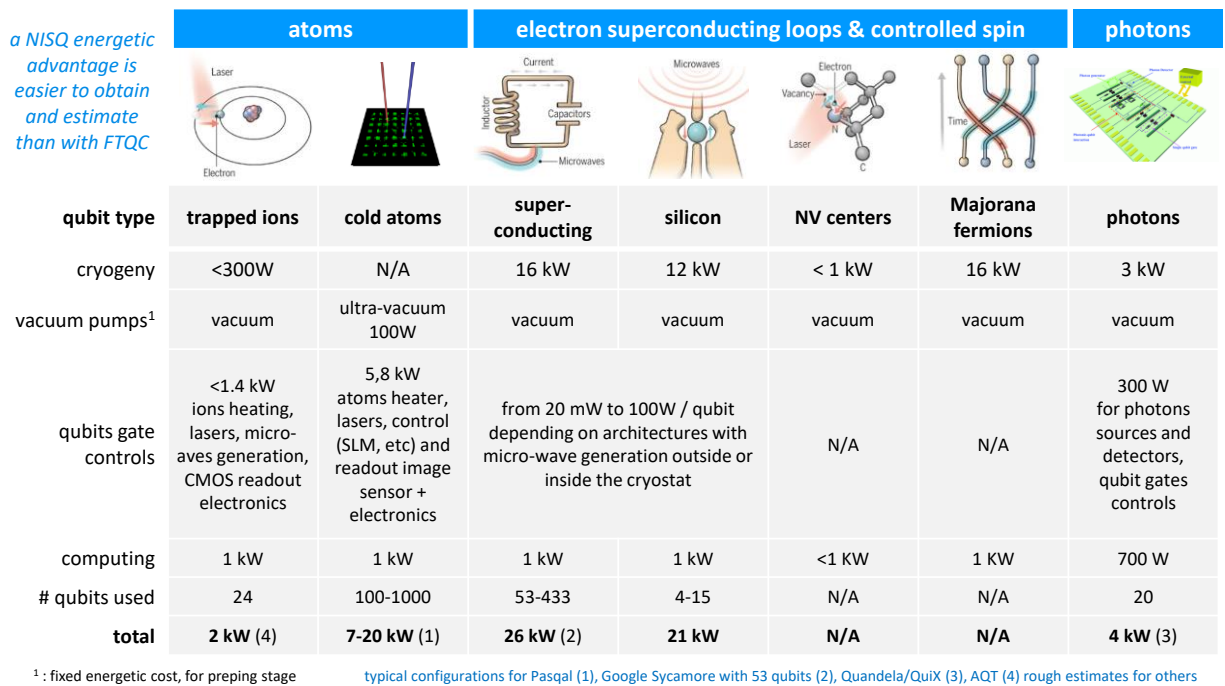

| *a NISQ energetic advantage is easier to obtain and estimate than with FTQC* | atoms | | electron superconducting loops & controlled spin | | | | photons |
|---|---|---|---|---|---|---|---|
| qubit type | trapped ions | cold atoms | super-conducting | silicon | NV centers | Majorana fermions | photons |
| cryogeny | <300W | N/A | 16 kW | 12 kW | < 1 kW | 16 kW | 3 kW |
| vacuum pumps¹ | vacuum | ultra-vacuum 100W | vacuum | vacuum | vacuum | vacuum | vacuum |
| qubits gate controls | <1.4 kW ions heating, lasers, micro-aves generation, CMOS readout electronics | 5,8 kW atoms heater, lasers, control (SLM, etc) and readout image sensor + electronics | from 20 mW to 100W / qubit depending on architectures with micro-wave generation outside or inside the cryostat | | N/A | N/A | 300 W for photons sources and detectors, qubit gates controls |
| computing | 1 kW | 1 kW | 1 kW | 1 kW | <1 kW | 1 kW | 700 W |
| # qubits used | 24 | 100-1000 | 53-433 | 4-15 | N/A | N/A | 20 |
| total | 2 kW (4) | 7-20 kW (1) | 26 kW (2) | 21 kW | N/A | N/A | 4 kW (3) |

¹ : fixed energetic cost, for preping stage          typical configurations for Pasqal (1), Google Sycamore with 53 qubits (2), Quandela/QuiX (3), AQT (4) rough estimates for others

Figure 30: existing QPU typical power drain and their source. Caveat: none of these systems provide a quantum advantage at this point in time (2023). Source: (cc) Olivier Ezratty, 2023.

| Brand | Existing commercial QPUs | Future NISQ regime QPUs |
|---|---|---|
| IBM | 127 qubits<br>Washington<br><50kW | 1,386 qubits<br>Flamingo<br><140 kW |
| D-Wave | 5,000 qubits<br>Advantage<br><30 kW | 7,000 qubits<br>Clarity<br><40kW |
| PASQAL | 100 atoms<br>Fresnel<br><3 kW | 300-1,000 atoms<br>Next gen<br><20 kW |
| QuEra | 256 atoms<br><3 kW | <20 kW |
| Quandela | 12 qubits<br>< 2 kW | TBD |

Figure 31: table comparing the power drain of existing QPUs and future NISQ-grade level QPUs from a few vendors. Source: Olivier Ezratty, consolidating vendor data and projections. If and when some of these future systems bring a quantum computing advantage in the near future, it could be also done with a related energetic advantage.

This is one of the goals and mission of the "Quantum Energy Initiative" launched in 2022 by Alexia Auffèves, Robert Whitney, Janine Splettstoesser and the author, and which advocates the creation of an interdisciplinary line of research around the energetics of quantum technologies[228]. The Quantum Energy Initiative also promotes a methodology defining clearly what is the energetic performance of a system and advocates for setting up full-stack modeling techniques to assess and optimize the energetics of quantum computers[229].

NISQ may land in a situation where its quantum advantage end-up being qualitative and energy related more than computing time related as shown in Figure 32. In a world of finite resources, this would make NISQ solutions totally relevant in the high-performance computing landscape.

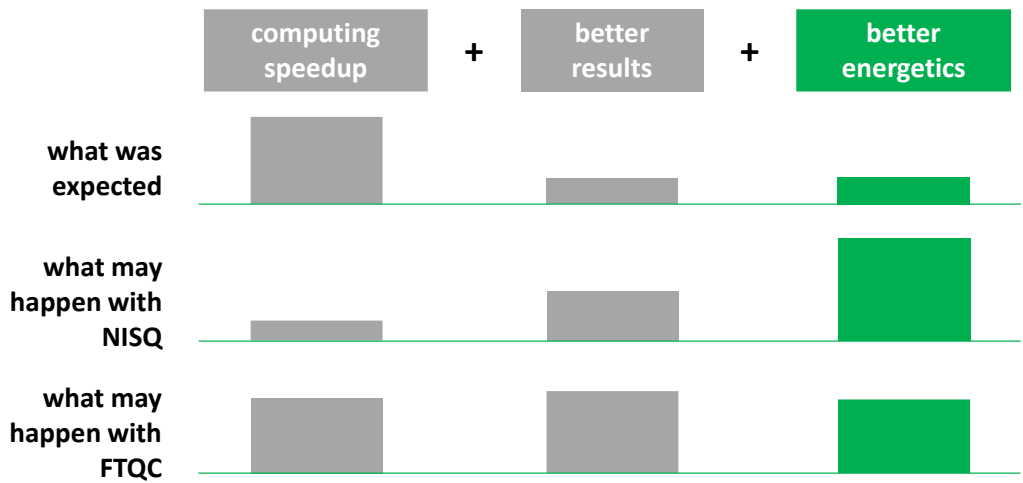

Figure 32: a new quantum advantages perspective. The bar height corresponds - without units - to the relative added value of the solution as compared to equivalent classical solutions. The prerequisite is of course that NISQ and FTQC algorithm bring some computing benefit or being at least be on par compared with classical computers achieving the same task. Source: (cc) Olivier Ezratty, 2023.

## V.   NISQ AND FTQC ROADMAPS

### Most impactful algorithms require FTQC

From a use case perspective, what is the difference between NISQ and FTQC?

We've seen that NISQ algorithms, when reaching a quantum advantage yet to be seen, cover a wide spectrum of optimizations, machine learning and physical simulations. Although it is not very well documented, their potential is moderate with the size of problems they could address.

Indeed, as we have seen in the previous parts of this paper, NISQ doesn't scale very well for at least three reasons: the difficulty to create very high-fidelity qubits that could enable mid-scale NISQ with several hundreds of qubits and gate cycles, the exponential scaling in the wrong direction of quantum error mitigation costs and totally unreasonable computing times, particularly with VQE algorithms used for various chemical simulations.

FTQC algorithms add several additional features:

- Solving problems with more variables like simulating larger molecules, solving larger combinatorial problems but deterministically, and larger quantum machine learning models.

- Various algorithms relying on a quantum Fourier transform like quantum phase estimates, quantum amplification estimates, HHL for linear algebra and partial differential equations. These are being used in quantum many-body simulation, quantum machine learning, financial applications and many other use cases. QPE (quantum phase estimate) based chemical simulation algorithms have computing times which scale slower than their NISQ VQE equivalent.

- Shor integer and discrete-log algorithms, whose main "business value" is definitively not in the "tech for good" domain, consisting in breaking secret keys in public key infrastructures as well as for symmetric keys sharing.

- Solving oracle-based search and optimization problems, using Grover algorithm and the likes. In some cases, it depends on the availability of various forms of quantum memory that are also yet to be seen. And it doesn't scale very well, bringing only a potential polynomial speedup.

The typical trouble with FTQC, these algorithms and their real world use cases is the sheer level of resources required in terms of physical qubits. Many papers have made such resource estimations, including Microsoft recent resource estimation tool already mentioned[26]. Also, like with NISQ algorithms like VQE, FTQC algorithms computing times may be prohibitive. It can be the case for both a Grover search algorithm or even a QPE algorithm to estimate various properties of a many-body quantum system[224].

- According to Xanadu and Volkswagen, simulating key properties of batteries (voltage, ionic mobility and thermal stability, including simulations of cathode materials using first-quantization algorithms) would require between 2,375 and 6,652 logical qubits with error rates $<10^{-12}$ as shown in Figure 33. Quantum computing time estimates are quite frightening. You quickly exceed one year of computing time, even with a QPU clock rate reaching 100 MHz. Right now, with superconducting qubits, we are in the 1.4 to 14 Khz clock range kHz[243].

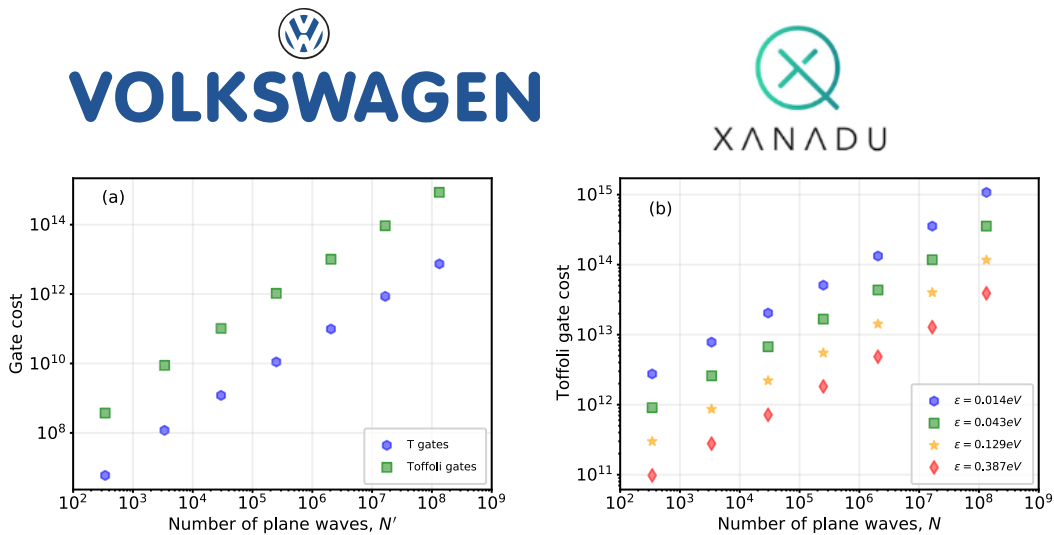

FIG. 11. **Non-Clifford gate cost for initial state preparation and quantum phase estimation.** (a) The non-Clifford gate cost due to Givens rotations used in the circuit for initial state preparation. (b) Toffoli gate cost of the quantum phase estimation algorithm. All calculations are done for the unit cell of $Li_2FeSiO_4$ with 156 electrons. The total number of qubits is 2,375 for $n_p = 4$ and 6,652 for $n_p = 9$. In the right figure we only depict Toffoli gate count, as the number of T gates is much smaller ($< 3 \times 10^6$). The total error $\varepsilon$ includes contributions from different approximations throughout the algorithm, but it does not take into account the error derived from a finite basis set. The slope of the Toffoli gate cost for fixed target precision is a consequence of the leading cost term in (100), $12\eta n_p \lceil (\pi\lambda)/(2\varepsilon_{QPE}) \rceil$, where $n_p = \lceil \log(N^{1/3}+1) \rceil$. These calculations were performed with the T-Fermion library [38].

Figure 33: simulating Li2FeSiO4 oxydes in batteries would cost over 6000 logical qubits. Source: Alain Delgado et al [230].

- In another source, PsiQuantum estimates that Li-ion battery simulation requires 16K logical qubits. Given PsiQuantum current plans, it would mean 160 million physical photonic qubits, as shown in Figure 35. Their paper provides estimates of the number of logical qubits and computing depth required for various molecular simulations and other algorithm needs. Part of the estimate logic is applicable to FBQC-based photonic qubits developed by PsiQuantum[231].

- According to Joonho Lee, simulating the FeMoCo molecular complex in nitrogenase simulation can be optimized using tensor hypercontraction. It would then require $\approx 2,142$ logical qubits, $3.2 \times 10^{10}$ Toffoli gates over 4 days run-time and with 4 million physical qubits[232], as shown in Figure 34. It is frequently touted as the key solution on the path to designing more efficient bioinspired fertilizers production methods. Practically speaking, getting some properties of FeMoCo is just the beginning of a more complex exploratory path to understanding how FeMoCo catalyzes the nitrogen ($N_2$) to ammonia ($NH_3$) conversion process in nitrogenase with its many molecular complexes and features[233][234].

| Algorithm | Reiher *et al.* FeMoCo [23] | | Li *et al.* FeMoCo [36] | |
|---|---|---|---|---|
| | Logical qubits | Toffoli count | Logical qubits | Toffoli count |
| Reiher *et al.* [23] (Trotter) | 111 | $5.0 \times 10^{13}$ | — | |
| Campbell and Kivlichan *et al.* [52,53] (qDRIFT) (D16), (D17) | 288 | $5.2 \times 10^{27}$ | 328 | $1.8 \times 10^{28}$ |
| qDRIFT with 95% confidence interval (D34) | 270 | $1.9 \times 10^{16}$ | 310 | $1.0 \times 10^{16}$ |
| Berry *et al.* [9] (single factorization) (B16), (B17) | 3,320 | $9.5 \times 10^{10}$ | 3,628 | $1.2 \times 10^{11}$ |
| Berry *et al.* [9] (sparse) (A17), (A18) | 2,190 | $8.8 \times 10^{10}$ | 2,489 | $4.4 \times 10^{10}$ |
| von Burg *et al.* [10] (double factorization) (C39), (C40) | 3,725 | $1.0 \times 10^{10}$ | 6,404 | $6.4 \times 10^{10}$ |
| This work (tensor hypercontraction) (44) (46) | 2,142 | $5.3 \times 10^9$ | 2,196 | $3.2 \times 10^{10}$ |

Figure 34: FeMoCo simulation requires at least 2000 logical qubits. Source: Joonho Lee et al [232].

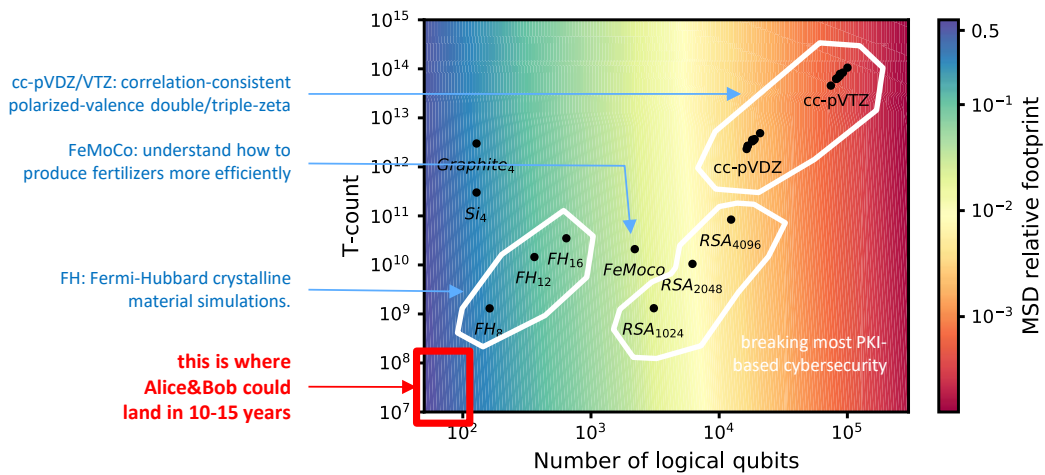

Figure 35: PsiQuantum estimates for implementing Fermi-Hubbard crystal material simulations, Shor algorithm on typical RSA key sizes and cc-pVDZ/VTZ molecular compounds. Source: Isaac H. Kim et al [231].

- Pricing derivatives would require 8K logical qubits, 54M T gates and a $\approx 10^{-8}$ logical qubit error rate[235].

- Option pricing using an amplitude estimation algorithm including an inverse QFT would require a computing depth between 3,927 and 285,204, assuming all-to-all qubits connectivity (see Figure 36). It provides a quadratic speed-up compared vs Monte Carlo classical simulations. It was tested at a very low scale on a 20-qubit IBM QPU[236].

| # | Single-qubit | CX | CCX | Depth |
|---|---|---|---|---|
| $m = 3$ | 2,091 | 2,056 | 90 | 3,927 |
| $m = 5$ | 12,768 | 9,078 | 378 | 17,332 |
| $m = 7$ | 52,275 | 37,132 | 1,530 | 70,916 |
| $m = 9$ | 210,144 | 149,290 | 6,138 | 285,204 |

Table 2: Single-qubit, CNOT, Toffoli gate counts and overall circuit depth required for the full amplitude estimation circuits for each instance in Fig. 8, as a function of the number of sampling qubits $m$. These figures assume all-to-all connectivity across qubits.

Figure 36: some resources requirements to implement a given option pricing algorithm. Source: Nikitas Stamatopoulos et al [236].

### In what order may NISQ and FTQC arrive?

John Preskill definition of NISQ implied that it was an intermediate path on the road to FTQC. One after the other. What if that sequence was not the only option? We've seen here how NISQ and FTQC were potentially two parallel routes with their own different tools and challenges, as summarized in Figure 37.

From the perspective of NISQ to FTQC, qubit fidelities and numbers must be improved to obtain some quantum advantage in the NISQ regime. If these qubits scale with good fidelities, we could quickly land in the FTQC regime. Indeed, when and if qubit fidelities reach the practical threshold for FTQC implementation, at around 99.9%, it will however be insufficient to implement NISQ over 100 qubits.

And we've seen that many NISQ algorithms require QPUs with fidelities way above 99.99% fidelities. This would mean that FTQC is a de facto more viable path to implementing even the so-called NISQ algorithms enabling some quantum advantage. This may explain why some physicists think that FTQC is the only viable path for obtaining a quantum advantage.

But you could also infer that it may be easier to create a few hundred high quality qubits for NISQ than a very large number of 99.9% well entangled qubits for FTQC. This is mandatory to obtain a real quantum advantage with NISQ QPUs since under 99.9% fidelities, QPUs are easy to emulate classically. If scaling qubits in the ten thousand to million number zone became impossible, it would mean that NISQ may be the only viable path. On the other hand, if we were able to build very high quality qubits and it scaled well, it could enable the creation of FTQC QPUs with a smaller number of physical qubits, reducing the scalability burden, particularly pertaining to cabling, control electronics and signals multiplexing.

| | NISQ | FTQC |
|---|---|---|
| physical qubits numbers | 50-1000s | 9000-millions |
| algorithmic qubit error rates $\epsilon$ | $10^{-3} \le \epsilon \le 10^{-7}$ | $10^{-5} \le \epsilon \le 10^{-15}$ |
| required physical qubit fidelities | 99.9% to 99.99999% | ≈99.9% |
| errors processing techniques | quantum error suppression | |
| | quantum error mitigation | quantum error correction |
| algorithms — VQE, QAOA, QML | Yes | Yes |
| algorithms — oracle based search | No | Yes |
| algorithms — QFT based | No | Yes (HHL, Shor, ...) |
| qubit challenges | average number of very high fidelity qubits | very large set of entangled qubits with good fidelities |
| other challenges | error mitigation scaling number of Pauli strings and shot with VQE | quantum memory / qRAM error correction overhead energetics |

Figure 37: some respective figures of merit and challenges of NISQ and FTQC. Source: (cc) Olivier Ezratty, 2023.

This remains an open question. How large can an entangled web of qubits be? Would it reach the famous quantum-classical bound? It deserves a better understanding of the "noise budget source" for various types of qubits[237]. Some consolidation would be welcomed, for example with superconducting qubits, on the scale of the noise sources between leakage, crosstalk, cosmic rays, control electronics jitter and the likes, and how far we could fight them[238]. In between, industry vendors like IBM are convinced that the line between NISQ and FTQC will blur, particularly with the help of various quantum error mitigation techniques.

The last option, which some pessimistic physicists consider to be the most plausible, is these paths are not viable. But you cannot prove that something is impossible to achieve, even if you ground your certainty in documented science. Human ingenuity is boundless. You can't determine its limits in advance.

There are also intermediate paths between NISQ and FTQC. One comes from Fujitsu, Osaka University and RIKEN in Japan and consists in reducing the number of physical qubits required to build logical qubits with using corrected and precise analog phase rotation gates involving a low overhead correction scheme instead of constructing it with costly combinations of error-corrected H and T gates[239]. This would enable the creation of useful early FTQC setups with only 10,000 physical qubits to support 64 logical qubits[240].

Another one is proposed by Quantinuum and involves a lightweight quantum error correction scheme adding a very low ancilla qubits overhead[241].

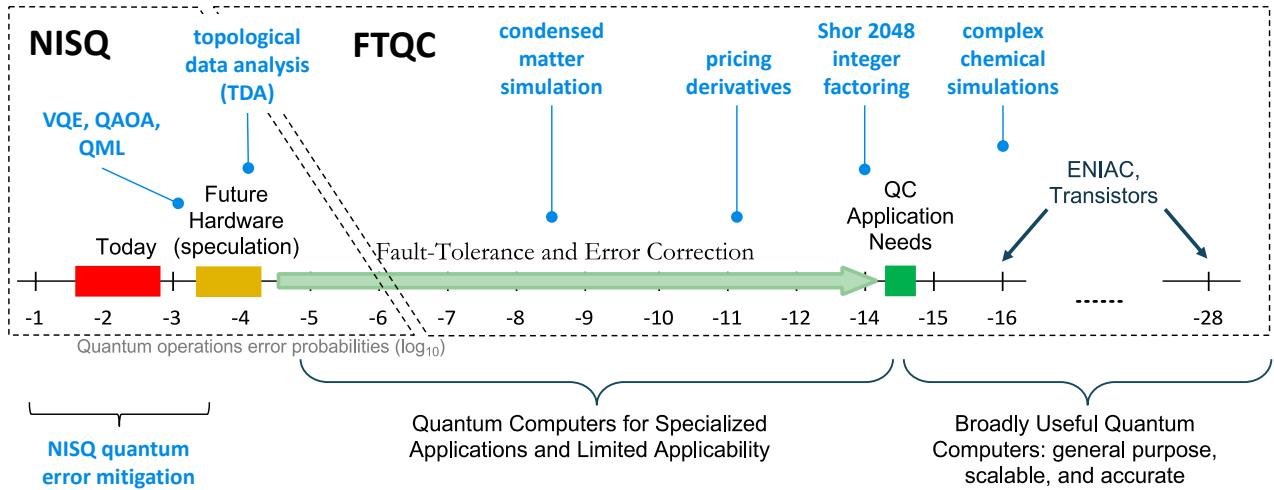

Figure 38: the road from NISQ to FTQC is uncertain. We could have a long NISQ road going through very high fidelity qubits and another path with FTQC logical qubits built with lesser quality qubits. All in all, one requirement is to be able to control entanglement of a very large number of quantum objects. ENIAC and transistors corresponds to the fidelity of triode (used in the late 1940s and early 1950s mainframe computers like the ENIAC and UNIVAC) and transistor logic operations in classical computing. Source: Bert de Jong[242] and Olivier Ezratty additions.

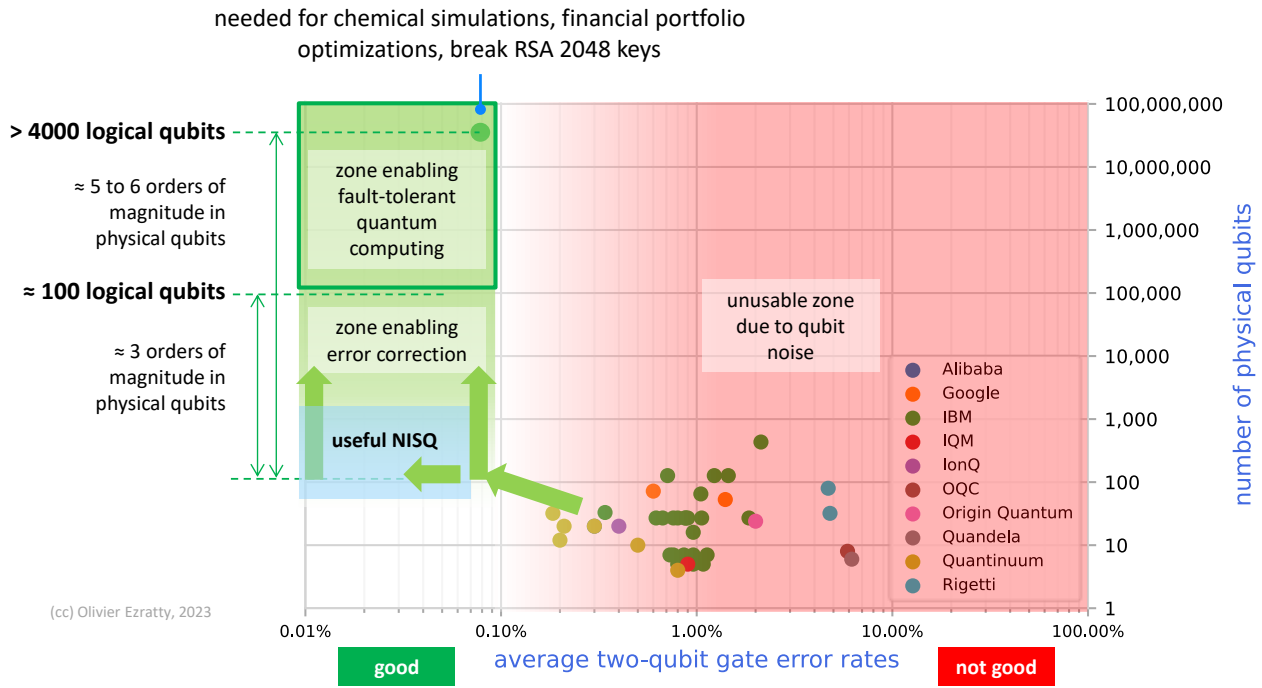

Figure 39: the paths to NISQ and FTQC are slightly different with an intermediate of very high quality qubits with NISQ and more lesser quality qubits for FTQC. Source: (cc) Olivier Ezratty, 2023, and vendor data.

## VI.   DISCUSSION

This paper highlighted many contradictions on the status of NISQ as a viable path to achieve some computational quantum advantage. It shows that it is linked to some lack of maturity of the technology but also to missing generic benchmarking techniques enabling multi-parameters performance and total cost of ownership comparisons between best-in-class classical and quantum computing solutions.

At this point in its development, the downsides of NISQ are manyfold:

- Practical NISQ is hard to achieve with existing hardware. It is a rather long-term goal in the roadmap of most hardware vendors.

- There are conflicting requirements for the number of qubits, fidelities and algorithms depth with existing and even prospective future hardware.

- NISQ algorithms designers do not investigate or document well enough how hardware resources requirements scale both in QPUs and in their classical part in order to reach some form of quantum advantage. This is a particularly hard task for heuristics based algorithms.

- Most QAOA and VQE algorithms do not scale well to a quantum advantage level with existing and near-term future hardware, particularly when you look at the details of their measurement steps that require at least a polynomial number of shots. On top of this, NISQ quantum advantage is highly use-case and algorithm dependent and not generic and in many cases, like with many-body simulations using VQE algorithms, computing times estimates are currently very high, up to largely exceeding a human lifetime. Many new theoretical bounds also show up that prevent NISQ scaling in a quantum advantage regime.

- Existing noisy NISQ gate-based algorithms actual implementations can most of the time be easily emulated on classical hardware. Shallow gate-based quantum algorithms can most of the time be efficiently emulated on classical computers using tensor network based techniques.

- Many useful quantum algorithms require FTQC hardware with millions if not billions of physical qubits.

- NISQ hardware vendors currently tend to oversell what can be done with their systems and fuel unjustified hype, mainly because they are raising funds and want to please potential investors in search of customers and short term revenue opportunities. Most hardware startups are still low TRL private research labs.

Taking the opposite stance and a longer term view, there are some potential upsides to turn NISQ into a practical reality although they all deserve some additional scrutiny:

- Short term quantum hardware may be entering the NISQ power range requirements in terms of qubit numbers and even fidelities, mainly from IBM (Heron processor, related to the 100x100 IBM challenge announced in November 2022[243]). Trapped ions may be just behind, but with strong scaling limitations and too slow gates.

- Many new quantum error mitigations techniques must be investigated and their benefit and overhead quantified. They can extend the reach of current and near term NISQ platform, although with their own scalability challenges. There may be a small range of quantum advantage potential in the small NISQ scale before quantum error mitigation reaches its limits.

- Analog quantum computing seems to be a more functional NISQ computing paradigm, despite its scaling capacity being unknown and probably limited due to the absence of error correction techniques. The related industry vendor space could extend well beyond the current neutral atoms offering, for example, with silicon qubits and trapped ions.

- NISQ algorithms developments indirectly drive a healthy competition between classical and quantum algorithms, which is likely to spur advancements in both areas.

- NISQ is a also learning path toward FTQC. Skipping NISQ to jump directly onto FTQC could be perceived as being a mistaken approach since failing with NISQ may also means directly failing with FTQC. But the NISQ route may be easier to take vs controlling millions of physical qubits with their huge scalability challenges, both at the quantum level (entanglement, fidelities) and classical level (cost of control, cooling). It's a trade-off between quality and quantity, as shown in Figure 40.

- NISQ systems could bring some quantum advantages with some algorithm quality advantage and an energetic advantage. It is still an uncharted territory to investigate.

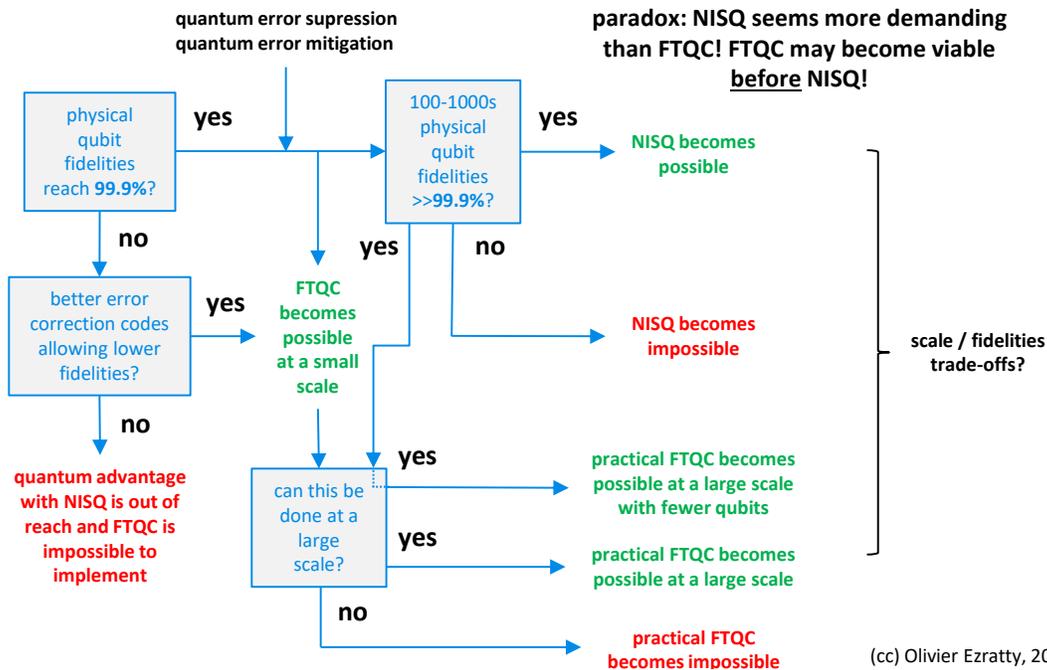

Figure 40: there are many scenarios for the advent of NISQ and FTQC QPUs. In one scenario, FTQC may become viable before NISQ. It is a matter of qubit fidelities threshold differences between the needs for FTQC and viable NISQ bringing some quantum advantage. But if NISQ is a path to create much higher fidelity qubits and it is possible to build them at scale, then NISQ could be the path to create FTQC QPUs with a smaller number of physical qubits per logical qubits. Source: (cc) Olivier Ezratty, 2023.

The tension between these downsides and cautious optimism is not just a "debate" on NISQ but is characteristic of an emerging field with blurry lines between fundamental research and vendors technology developments and their commercialization. My intent here was also to showcase the enormous gap between the scientific and technological reality of quantum computing and the current overpromises coming from some analysts and industry vendors. The current abusive buzz on the so-called business readiness of quantum computing could seriously backfire with unintended negative consequences[244].

Quantum computing is a rather long term quest and should be understood as such, particularly by governments, policy makers and investors. It shouldn't however prevent corporations from investigating the whereabouts of quantum computing, to learn about it, and to evaluate early stage algorithms and hardware solutions, particularly in the analog quantum computing space. It can help them reassess their large scale computing needs, their unaddressed complex business problems and drive some healthy emulation with classical computing specialists.

The paper also illustrates how a journey in quantum computing is highly cross-discipline and why more connections and common understanding must be developed between quantum computers scientists and technology developers, quantum algorithms and software developers, and their counterparts in classical computing.

The author warmly thanks David Amaro (Quantinuum), Alain Chancé (Molket)[245], Pierre Desjardins (C12 Quantum Technologies), Vincent Elfving (Pasqal), Marco Fellous-Asiani (Centre of New Technologies University of Warsaw), Loïc Henriet (Pasqal), Michel Kurek (Multiverse), Jean-Baptiste Latre (Qualitative Computing), Mark Mattingley-Scott (Quantum Brilliance), Joseph Mikael (EDF), Stéphane Requena (HQI/GENCI), Dario Rosa (IBS Korea), Aritra Sarkar (QuTech), Jean Senellart (Quandela), Simone Severini (AWS), Robert Whitney (CNRS LPMMC, QEI), Xavier Waintal (CEA IRIG) and Raja Yehia (ICFO) for their feedback on the paper (which doesn't mean an endorsement of all its content) and recognizes useful discussions with and insights from Alexia Auffèves (CNRS MajuLab, CQT, QEI), Cyrille Allouche (Eviden (Atos)), Thomas Ayral (Eviden (Atos)), Jerry Chow (IBM), Jay Gambetta (IBM), Cornelius Hempel (PSI), John Preskill (Caltech), and Pierre Perrot (Eureptech).

(cc) Olivier Ezratty, written between February and May 2023. V1 published on arXiv on May 17th, V2 on May 22st and V3 on June 13th.

# VII.    SOURCES AND NOTES